\documentclass[preprint]{JHEP} 
\usepackage{amsmath}
\usepackage{amssymb,amsfonts}
\usepackage{epsfig}

\newcommand\fverb{\setbox\pippobox=\hbox\bgroup\verb}
\newcommand\fverbdo{\egroup\medskip\noindent%
            \fbox{\unhbox\pippobox}\ }
\newcommand\fverbit{\egroup\item[\fbox{\unhbox\pippobox}]}
\newbox\pippobox

\newcommand{\eqa}[1]{\begin{eqnarray*} #1 \end{eqnarray*} }

\newcommand{\nn}{\nonumber}

\newcommand{\und}[1]{\noindent \underline{ #1} }

\def\co{: \;\;\;}

\def\rma{r_{\rm max}}
\newcommand{\be}{\begin{equation}}
\newcommand{\ee}{\end{equation}}
\newcommand{\ben}{\begin{enumerate}}
\newcommand{\een}{\end{enumerate}}
\renewcommand{\sp}{\ ,\qquad}
\makeatletter
\renewcommand{\@makefnmark}{\mbox{$^{\ddagger\@thefnmark}$}}
\renewcommand{\subsection}{\@startsection
  {subsection}{2}{0pt
}{-\baselineskip}{0.5\baselineskip}
  {\normalfont\normalsize\itshape}}
\makeatother
\numberwithin{table}{section}

\title{Thermodynamics of Spinning Branes and their Dual Field Theories}

\author{T. Harmark and N.A. Obers\thanks{
Work supported in part by TMR network ERBFMRXCT96-0045.}\\
    Niels Bohr Institute and Nordita, Blegdamsvej 17, DK-2100 Copenhagen, Denmark \\
    E-mail: \email{harmark@nbi.dk}, \email{obers@nordita.dk} }

\preprint{ NBI-HE-99-37  \\
NORDITA-1999/61 HE \\
\hepth{9910036}}    

\abstract{We present a general analysis of the thermodynamics of
spinning black $p$-branes of string and M-theory. This is
carried out both for the asymptotically-flat and near-horizon
case, with emphasis on the latter. In particular, we use the
conjectured correspondence between the near-horizon brane
solutions and field theories with 16 supercharges in various
dimensions to describe the thermodynamic behavior of these field
theories in the presence of voltages under the R-symmetry.
Boundaries of stability are computed for all spinning branes both
in the grand canonical and canonical ensemble, and the effect of
multiple angular momenta is considered. A recently proposed
regularization of the field theory is used to compute the
corresponding boundaries of stability at weak coupling. 
For the D2, D3, D4, M2 and M5-branes
the critical values of $\Omega/T$ in the weak and strong coupling
limit are remarkably close. Finally, we also show that for the spinning
D3-brane the tree level $R^4$ correction supports the conjecture
of a smooth interpolating function between the free energy at weak
and strong coupling.}


\begin{document}

\maketitle 

\section{Introduction and conclusion}

In the early 1970s, two important discoveries were made which have
played a dominant role in theoretical physics ever since. The
first discovery, by Bekenstein \cite{Bekenstein:1973ur} and
Hawking \cite{Hawking:1974df}, was that four-dimensional black
holes have thermodynamic properties due to Hawking radiation.
Thus, by studying thermodynamics of black holes one probes the
nature of quantum gravity. In the framework of string and
M-theory, this discovery has been one of the main motivations to
consider the thermodynamics of black $p$-branes
\cite{Horowitz:1991cd,Gubser:1996de,Klebanov:1996un}. The second
discovery, by 't Hooft \cite{'tHooft:1974jz}, was that non-Abelian
gauge theories simplify in the 't Hooft limit. In this limit the
planar diagrams dominate and the theories thus become more
tractable.

More recently, it has become clear that these two discoveries are
in fact connected through the  conjectured correspondence between
the near-horizon limit of brane solutions in string/M-theory and certain
quantum field theories in the large $N$ limit
\cite{Maldacena:1997re,Itzhaki:1998dd}. As
a consequence of this correspondence, studying the thermodynamic
properties of black $p$-branes not only probes quantum
gravity, but can in addition provide information about the
thermodynamics of quantum field theories in the large $N$ limit.

In particular, for the non-dilatonic branes (D3,M2,M5) the
near-horizon limit of the supergravity solutions
 has been conjectured \cite{Maldacena:1997re}
 to be dual to a certain limit of the
 corresponding conformal field theories (see also
 Refs. \cite{Gubser:1998bc,Witten:1998qj} for an elaboration of the
conjecture at the level of the partition function and correlation functions).
 In these AdS/CFT
 correspondences, the
near-horizon background geometry is of the form $AdS_{p+2} \times
S^{d-1}$ and the dual field theories are conformal. Moreover, for
the more general dilatonic branes of type II string theory
preserving 16 supersymmetries, similar duality relations have also
been obtained \cite{Itzhaki:1998dd}, which may be characterized
more generally as Domain Wall/QFT correspondences
\cite{Boonstra:1998mp,Behrndt:1999mk}. See Ref.
\cite{Aharony:1999ti} for a comprehensive review and Refs.
\cite{Klebanov:1999ku,Petersen:1999zh,DiVecchia:1999yr} for some
introductory lectures on the AdS/CFT correspondence and field
theories in the large $N$ limit.

A common feature of these dualities between near-horizon
backgrounds and field theories, is that the supergravity black
$p$-brane solution exhibits an $SO(d)$ isometry (where $d=D-p-1$
is the dimension of the transverse space) which manifests itself
as the R-symmetry of the dual field theory. As a consequence, by
considering black $p$-brane solutions that rotate in the
transverse space, we expect on the one hand to learn more about
the field theory side, and on the other hand, be able to perform
further non-trivial tests of the duality conjectures that include
the dependence on this R-symmetry group. In particular, as will be
reviewed below, the thermodynamics on the two sides provides a
useful starting point for such a comparison.

The first construction of spinning branes solutions, rotating
in the transverse space,  can be found in
 Refs. \cite{Horowitz:1996tm,Cvetic:1996xz,Cvetic:1996dt,Cvetic:1996ek}
from which, in principle the most general black $p$-brane solution
 \cite{Cvetic:1999xp} can be derived by oxidization.
Also, various spinning brane solutions
\cite{Russo:1998mm,Csaki:1998cb,Kraus:1998hv,Russo:1998by,Csaki:1999vb,Correia:1999bt,Brandhuber:1999jr}
have recently been constructed and employed with the purpose
to  provide extra dimensionfull parameters in the
decoupling of the unwanted KK modes in the context of obtaining
QCD  in various dimensions via the AdS/CFT correspondence
\cite{Witten:1998zw}. Other examples of spinning brane solutions
include the spinning NS5-brane \cite{Sfetsos:1999pq} and
rotating Kaluza Klein black holes \cite{Larsen:1999pp}.
 Spinning branes have
also been used \cite{Cai:1999ad} in the study of D-brane probes
\cite{Tseytlin:1998cq,Kiritsis:1999ke,Kiritsis:1999tx}. Many
aspects of the case in which the rotation does not lie in the
transverse space
\cite{Hawking:1998ct,Hawking:1998kw,Berman:1999mh,Caldarelli:1999xj,Hawking:1999dp},
generally referred to as the Kerr-AdS type, have also been
considered in view of the AdS/CFT correspondence but will not be
considered in this paper.

It is interesting in its own right to study  the thermodynamic
properties of black $p$-branes and rotating versions thereof,
since this may teach us more about black brane physics.
In particular, the near-horizon solution is thermodynamically much better
behaved than the asymptotically-flat solution,
which along with its relevance  to a certain limit of the dual
field theories, makes it very interesting to study the
thermodynamics in this case. For the non-dilatonic branes the study
of the thermodynamic stability has been initiated\footnote{Other aspects of the
thermodynamics in relation to the AdS/CFT correspondence and
holography were studied e.g. in Refs.
\cite{Barbon:1998ix,Barbon:1998cr,Chamblin:1999tk,Chamblin:1999hg,Martinec:1999bf}.}
in a number of recent papers
\cite{Gubser:1998jb,Cai:1998ji,Cvetic:1999ne,Cai:1999hg,Cvetic:1999rb}.
The stability for the D3-brane with one non-zero angular momentum
was addressed in \cite{Gubser:1998jb} followed by an
analysis both in the grand canonical and canonical ensemble for
all non-dilatonic branes \cite{Cai:1998ji}. It was found that
these two ensembles are not equivalent. An analysis of the
critical behavior near these boundaries was also performed and
shown to obey scaling laws of statistical physics. The case of
multiple angular momenta was considered in Ref.
\cite{Cvetic:1999rb} for both ensembles.

 Furthermore, in order to
compare with the field theories, a regularization method
\cite{Gubser:1998jb,Cvetic:1999rb} has been proposed and used in
order to compare the stability behavior obtained from the
supergravity solution with that of the corresponding field theory
in the weakly coupled limit. The angular momenta take
values in the isometry group of the sphere, and hence map onto the R-charges
in the dual field theory. Therefore, the angular velocities on the
brane correspond in the field theory to voltages under the R-symmetry.
In the presence of these voltages, a regularization is required, since
for massless bosons with non-zero R-charge negative thermal occupation
numbers occur. For the D3-brane case, it was found that the regulated
field theory analysis predicts a similar upper bound on the
angular momentum (or R-charge) density as obtained from the
near-horizon brane solution. The critical exponents obtained from
the supergravity solution are, however, not reproduced, though a
mean field theory analysis has been suggested to cure this
discrepancy. Finally, Ref. \cite{Cvetic:1999rb} also presents
evidence for localization of angular momentum on the brane
outside the region of stability, and the occurrence of  a first
order phase transition.

The comparison of boundaries of stability in the two dual sides is
one way to obtain evidence and predictions of the correspondence
between near-horizon brane solutions and field theories. Another
route, that also uses thermodynamic quantities is consideration of
the free energy which, in the non-rotating case,
has been computed  from the Euclidean action
by a suitable regularization method
\cite{Hawking:1983dh,Witten:1998zw}. For non-rotating D3, M2 and M5-branes
it has been conjectured that there exists a smooth interpolating function
connecting the two limits \cite{Gubser:1998nz}.
In particular for the D3-brane one finds that for the
near-horizon $AdS_5 \times S^5$ limit the free energy differs by a
factor 3/4 from the weakly coupled N=4 SYM expression. Since the
former limit corresponds to the strong 't Hooft coupling limit, it
can  be envisaged that higher derivative string corrections on the
supergravity modify this result in such a way that minus the free energy
increases towards the weak coupling limit. This conjecture was
tested \cite{Gubser:1998nz,Pawelczyk:1998pb} by computing the
correction to the free energy arising from the tree level $R^4$
term in the type IIB effective action, and shown to be in
agreement. In this spirit, the study of such corrections is also
interesting to perform in the presence of rotation.

In this paper, we will address various issues related to  the
developments described above, with emphasis on a general treatment
for all black $p$-branes that are 1/2 BPS solutions of string and
M-theory in the extremal and non-rotating limit. This includes the
M2 and M5-branes of M-theory and the D and NS-branes of
string theory. We will first write down the general
asymptotically-flat solution
of these spinning black $p$-brane in $D$ dimensions.
Since the transverse space is $d = D-p-1$ dimensional, these
spinning solutions are characterized by a set of angular momenta
 $l_i$, $i=1 \ldots n$ where $n={\rm rank} (SO(d))$,
 along with the non-extremality parameter $r_0$ and another
parameter $\alpha$ related to the charge. Using standard methods of
black hole thermodynamics we compute the relevant thermodynamic
quantities of the general solution and show that the conventional
Smarr formula is obeyed. (see Section \ref{secrotbra}).

Our main interest, however, will be in the near-horizon limit of
these spinning branes, which we will also compute in generality.
The corresponding thermodynamics that results in this limit will
also be obtained. In the near-horizon limit the charge and chemical
potential become
constant and are not thermodynamic parameters anymore, so that the
thermodynamic quantities are given in terms of the $n+1$
supergravity  parameters $(r_0,l_i)$. In particular, we derive
and check a modified Smarr law for the near-horizon background which
is due to a different scaling of the solution as compared to  the
asymptotically-flat case. One also finds a simple formula for the
Gibbs free energy for any near-horizon spinning black $p$-brane
solution with $d$ transverse dimensions
\begin{equation}
\label{gibbs}
 F = - \frac{V_p V(S^{d-1})}{16\pi G} \frac{d-4}{2}
r_0^{d-2}
\end{equation}
In a low angular momentum expansion, we rewrite this expression in
terms of the intensive thermodynamic quantities, the temperature
$T$ and the angular velocities $\Omega_i$.
 For comparison, we then use the correspondence with field
theory in the large  $N$ limit (and  appropriate limit of the 't
Hooft coupling limit in the case of D-branes), to write this free
energy in terms of the field theory variables. (see Section
\ref{secnhlimit}).

We proceed with presenting a general analysis of the boundaries of
stability in both the grand canonical ensemble (with thermodynamic
variables $(T,\Omega_i)$) and the canonical ensemble (with
thermodynamic variables $(T,J_i)$) of the near-horizon spinning
branes. While there is a one-to-one correspondence between the
$n+1$ supergravity variables and the extensive quantities $(S,J_i)$,
the map to the intensive ones $(T,\Omega_i)$ or the mixed
combination $(T,J_i)$ involves a non-invertible function, a fact which is
crucial to  the stability analysis. We will show in particular
that for general $d$, the two ensembles are not equivalent and
that increasing the number of equal-valued angular momenta
enlarges the
stable region\footnote{Except for the cases $d=8,9$, which have
no stability boundary for one non-zero angular momentum.}. For
one non-zero angular momentum we find, in the grand canonical
ensemble, that the region of stability (for $d \geq 5$) is
determined by the condition
\begin{equation}
J \leq \sqrt{\frac{d-2}{d-4}} \frac{S}{2 \pi}
\end{equation}
so that there is an upper bound on the amount of angular momentum
the brane can carry in order to be stable. Put another way, at a
critical value of the angular momentum density (which equals the
R-charge density in the dual field theory) a phase transition
occurs. The supergravity description also determines
an upper bound on the angular velocity,
\begin{equation}
\Omega \leq \frac{2 \pi}{ \sqrt{(d-2)(d-4)}} T
\end{equation}
which is saturated at the critical value of the angular momentum.
As a byproduct of the analysis we obtain an exact
expression of the Gibbs free energy in terms of $(T,\Omega)$ for
all branes in the case of one non-zero angular momentum\footnote{
 The case $d=4$, which can be seen from
\eqref{gibbs} to be special since the free energy vanishes, will
be treated separately. In this case, the temperature and angular
velocity are not independent, so that the phase diagram is degenerate.}.
 We will also comment on the nature of
the instability and discuss the setup to be solved in order to
determine whether there is some region of parameter space in which
phase mixing is thermodynamically favored, so that the angular
momentum localizes on the brane. Finally, we give a uniform
treatment of the critical exponents for all spinning branes in
both ensembles and show that all of these are 1/2, a value
which satisfies scaling laws in statistical physics. (see Section
\ref{seccritbeh}).

An important question is to what extent do we observe the above
stability phenomena in the large $N$ limit of the dual field
theory, also at weak coupling. To this end, extending the method
of Ref. \cite{Gubser:1998jb}, we obtain in an ideal gas
approximation the free energies of the field theories for the
case of the M-branes and the D-branes of type II string theory.
We review and extend the interpolation
conjecture, stating that the free energy smoothly interpolates
between the weak and strong coupling limit.
The free energies in
the weakly coupled regime enable us to corroborate these
conjectures by computing the boundaries of stability in this
regime. The corresponding critical values of the dimensionless
quantity $\Omega/T$  for the D2, D3, D4, M2 and M5-branes 
are
remarkably close in the weak and strong coupling limit. (see
Section \ref{secfieldtheory}). 

Finally, we also test the interpolation conjecture by considering
the free energy. We first establish that for all near-horizon spinning branes 
the on-shell Euclidean action 
reproduces the thermodynamically obtained Gibbs free energy
\eqref{gibbs}.  For the spinning D3-brane
we then calculate, in a weak angular momentum expansion, the
correction to the free energy due to the tree-level $R^4$ term in
the type IIB effective action. This order $\lambda^{-3/2}$
correction is positive   (in the range of validity) and hence
supports the conjectured existence of a smooth interpolating
function between the free energy in the weak and strong coupling
limit. (see Section \ref{secfefromaction}). 

A number of appendices are included: Appendix \ref{appenergy}
gives a  general discussion of (non-rotating) black $p$-branes,
including those that preserve a lower amount of supersymmetry. We
also find the thermodynamic quantities and Smarr formula for both
the asymptotically-flat and near-horizon solutions. Appendix
\ref{sphcoor} reviews spheroidal coordinates which are relevant
for the explicit form of spinning brane backgrounds. Appendix
\ref{appeuclnhsol} shows how the Euclidean spinning brane solution
can be obtained from the Minkowskian solution. Appendix
\ref{basisch} discusses the change of variables from the
supergravity variables to the intensive thermodynamic variables in
a weak angular momentum expansion. Finally, Appendix \ref{apppoly}
gives various useful expressions for the polylogarithms which are
used in Section \ref{secfieldtheory} to compute  the free energies
of the weakly coupled field theories in the presence of voltage
under the R-symmetry.


\newcommand{\Ord}{{\cal{O}}}
\newcommand{\eul}{{\tilde{l}}}

\section{General spinning p-branes
\label{secrotbra} }

More than forty years after the discovery of the Schwarzschild
black hole metric, Kerr presented in 1963 the first metric for a
rotating black hole \cite{Kerr:1963ud}. About twenty years later,
this was generalized to neutral rotating black holes of arbitrary
dimensions in \cite{Myers:1986un}. In
\cite{Horowitz:1996tm,Cvetic:1996xz,Cvetic:1996dt} these were
further generalized to charged rotating black hole solutions of
the low-energy effective action of toroidally compactified string
theory. In \cite{Cvetic:1996ek} the first spinning brane solutions
appeared and recently spinning brane solutions of type II
string theory and M-theory have been presented in
\cite{Russo:1998mm,Csaki:1998cb,Kraus:1998hv,Sfetsos:1999pq,Cvetic:1999xp}.

In this section we consider the general spinning brane solutions
of string theory and M-theory. The general solution is presented
in Section \ref{subsecspinsol} and in Section \ref{secthermdyn}
we derive the thermodynamic quantities of the general spinning
brane solutions.

\subsection{The spinning black $p$-brane solutions
\label{subsecspinsol} }

In this section we present the general charged spinning black
$p$-brane solution with a maximal number of angular momenta for
branes of string theory and M-theory. These solutions have the
property that they are 1/2 BPS states in the extremal and
non-rotating limit. Thus, they include the D- and NS-branes of 10-dimensional
string theory\footnote{Note that the D-branes of type I string
theory and the NS-branes of heterotic string theory are included
in this class of branes. When discussing the dual field theories
in the near-horizon limit we restrict to type II string theory and
M-theory only.}
 and the M-branes of 11-dimensional M-theory, as well as the
branes living in toroidal compactifications of these theories. The
solutions can be derived by oxidizing spinning charged black hole
solutions in a $D-p$ dimensional space-time
\cite{Horowitz:1996tm,Cvetic:1996xz,Cvetic:1996dt,Cvetic:1996ek,Russo:1998mm,Csaki:1998cb,Kraus:1998hv,Sfetsos:1999pq,Cvetic:1999xp}.

We only write the solutions for electric branes, since the
magnetic solutions can easily be obtained by the standard
electromagnetic duality transformation. In our conventions the
coordinate system is taken to be $(t,y^i,x^a)$, where $t$ is the
time, $y^i$, $i=1\ldots p$ the spatial world-volume coordinates
and $x^a$ the transverse coordinates. The space-time dimension is
denoted by $D$, so that $d = D - p - 1$ is the dimension of the
transverse space. The spinning brane solutions given below are
solutions of the action
\begin{equation}
\label{pbact}
I = \frac{1}{16\pi G} \int d^D x \sqrt{g} \Big( R
- \frac{1}{2} \partial_\mu \phi \partial^\mu \phi
- \frac{1}{2 (p+2)!} e^{a\phi} F_{p+2}^2 \Big)
\end{equation}
where \( F_{p+2} \) is the $(p+2)$-form electric field strength.
This action arises as  part of the 10-dimensional string effective
action in the Einstein frame or the 11-dimensional supergravity
action, and toroidal compactifications of these
actions\footnote{Toroidal compactifications of the supergravity
introduce more scalars in addition to the dilaton, which can be
ignored since these moduli do not affect the background solution
and its resulting thermodynamics.}. The value of $a$ is a
characteristic number for each brane, and for branes that are 1/2
BPS in the  extremal and non-rotating limit, one has the relation
\begin{equation}
\label{aeq}
2(D-2) = (p+1)(d-2) + \frac{1}{2}a^2 (D-2)
\end{equation}
Appendix \ref{appenergy} reviews\footnote{This appendix also
gives the general thermodynamic relations for all non-rotating
black $p$-branes, including those which preserve less than half of
the supersymmetries.} more general black brane solutions that do
not fulfill this identity and preserve a smaller amount of
supersymmetry (See also Table \ref{tabbranes} for the values of
$a$ for each of the branes that we consider).

As described in Appendix \ref{sphcoor}, the spinning solutions
depend on a set of angular momentum parameters \( l_1,l_2,...,l_n
\) where $n= [\frac{d}{2}]$ is the rank of $SO(d)$. Two further
parameters that characterize the solutions are the non-extremality
parameter $r_0$ and a dimensionless parameter $\alpha$, related to
the charge. In particular, $r_0=0$ corresponds to the extremal
$p$-brane solution, while $\alpha =0$ corresponds to a neutral
brane. The relation of these parameters to the thermodynamic
quantities of the solution will be discussed in Section
\ref{secthermdyn}. We restrict ourselves to the cases for which
the transverse dimensions lie in the range $ 3 \leq d \leq 9$ so
that the brane solutions are asymptotically flat.

The metric of a charged spinning $p$-brane solution of the action
\eqref{pbact} then takes the form
\begin{eqnarray}
\label{solmet}
ds^2 &=& H^{-\frac{d-2}{D-2}}
\Big( - f dt^2 + \sum_{i=1}^p (dy^i)^2 \Big)
+ H^{\frac{p+1}{D-2}} \Big( \bar{f}^{-1} K_d dr^2
+ \Lambda_{\alpha \beta} d\eta^\alpha d\eta^\beta \Big)
\nn \\ &&
+ H^{-\frac{d-2}{D-2}} \frac{1}{K_d L_d} \frac{r_0^{d-2}}{r^{d-2}}
\Big( \sum_{i,j=1}^n l_i l_j \mu_i^2 \mu_j^2 d\phi_i d\phi_j
- 2 \cosh \alpha \sum_{i=1}^n l_i \mu_i^2 dt d\phi_i \Big)
\end{eqnarray}
The electric dilaton is
\begin{equation}
\label{soldil}
e^\phi = H^{\frac{a}{2}}
\end{equation}
and the electric potential $A_{p+1}$ (with field strength $F_{p+2}=dA_{p+1}$)
is given by
\begin{equation}
\label{solpot} A_{p+1} = (-1)^p \frac{1}{\sinh \alpha} \Big( H^{-1}
-1 \Big) \Big( \cosh \alpha dt - \sum_{i=1}^n l_i \mu_i^2 d\phi_i
\Big) \wedge dy^1  \wedge \cdots \wedge dy^p
\end{equation}
Here, we have used spheroidal coordinates for the flat transverse space
metric
\begin{equation}
\label{flattr}
\sum_{a=1}^d (dx^a)^2
= K_d dr^2 + \Lambda_{\alpha \beta} d\eta^\alpha d\eta^\beta
\end{equation}
the explicit form of which can be found in Appendix \ref{sphcoor},
which also gives the angular dependence of the quantities $\mu_i$.
Moreover, we have defined
\begin{subequations}
\label{defs}
\begin{equation}
\label{Ldeq}
L_d = \prod_{i=1}^n \Big( 1 + \frac{l_i^2}{r^2} \Big)
\sp
H = 1 + \frac{1}{K_d L_d} \frac{r_0^{d-2} \sinh^2 \alpha}{r^{d-2}}
\end{equation}
\begin{equation}
f = 1 - \frac{1}{K_d L_d} \frac{r_0^{d-2}}{r^{d-2}} \sp
\bar{f} = 1 - \frac{1}{L_d} \frac{r_0^{d-2}}{r^{d-2}}
\end{equation}
\end{subequations}
and we note that the harmonic function $H$ is such that the branes
are asymptotically flat.

 The physical situation that this solution describes is
a charged black $p$-brane rotating in the angles
$\phi_1,\phi_2,...,\phi_n$ (see Appendix \ref{sphcoor}). The
rotation is static, meaning that the points of the $p$-brane move
with time, but that the total set of points of the brane in the
embedding space does not change with time. Thus, the solution
describes a spinning charged black $p$-brane.

\subsection{Thermodynamics of spinning branes
\label{secthermdyn} }

We proceed with describing some general physical properties of the
solution given in (\ref{solmet}), (\ref{soldil}) and
(\ref{solpot}), and the computation of its relevant thermodynamic
quantities.

The horizon is at $r=r_H$ where $r_H$ is the highest root of the
equation $\bar{f} (r) = 0$, so that
\begin{equation}
\label{horizon}
L_d (r_H) r_H^{d-2} = r_0^{d-2}
\end{equation}
where $L_d$ is defined in \eqref{Ldeq}. On the other hand, the
solution of the equation \( f (r) = 0 \) with the maximal possible
value of $r$ describes the so-called ergosphere, which coincides
with the horizon for special values of the angles \( \theta \) and
\( \psi_1,\psi_2,...,\psi_{d-n-2} \). It is useful to find a
coordinate transformation to a system in which these two
hyper-surfaces coincide. To this end we write
\begin{equation}
\label{newcoor} \tilde{t} = t \sp  \tilde{\phi}_i = \phi_i -
\Omega_i t \sp  i=1 \ldots n
\end{equation}
with all other coordinates unchanged. Thus, we want to find
$ \{\Omega_i,i=1\ldots n \} $  so that
\begin{equation}
\label{eqkill}
g_{\tilde{t} \tilde{t}} \Big|_{r=r_H} = 0
\end{equation}
In the transformed frame one has
\begin{equation}
\label{newgtt}  g_{\tilde{t}\tilde{t}} = g_{tt} + \sum_{i,j=1}^n
\Omega_i \Omega_j g_{\phi_i \phi_j} + 2 \sum_{i=1}^n \Omega_i g_{t
\phi_i}
\end{equation}
so that (\ref{eqkill}) can be written as
\begin{equation}
\label{eqkill2}
g_{tt}\Big|_{r=r_H}
+ \sum_{i,j=1}^n \Omega_i \Omega_j g_{\phi_i \phi_j}\Big|_{r=r_H}
+ 2 \sum_{i=1}^n \Omega_i g_{t \phi_i}\Big|_{r=r_H} = 0
\end{equation}
Since Eq. (\ref{eqkill2}) should hold for all angles, we can
consider the special choice  for which $\mu_i = 1$ and $\mu_{j\neq
i} = 0$. Then (\ref{eqkill2}) becomes
\begin{equation}
\label{eqkill3} g_{tt}\Big|_{r=r_H} + \Omega_i^2 g_{\phi_i
\phi_j}\Big|_{r=r_H} + 2 \Omega_i g_{t \phi_i}\Big|_{r=r_H} = 0
\sp i = 1 \ldots n
\end{equation}
which is satisfied by
\begin{equation}
\Omega_i = \frac{l_i}{(l_i^2 + r_H^2) \cosh \alpha} \sp  i=1 \ldots
n
\end{equation}
The new coordinate system defined in (\ref{newcoor}) can be seen
as comoving coordinates on the horizon, i.e. coordinates for which
the points on the brane in the embedding space do not move with
time\footnote{In Ref. \cite{Youm:1997rz} it was argued
that the comoving frame is the natural frame for studying
       thermodynamics of rotating black holes and that the
statistical analysis of
rotating black holes is simplified in this frame.}.
 From the definition (\ref{newcoor}) it then follows that \(
\Omega_i \)  is the angular velocity of a particle on the horizon
with respect to the angle \( \phi_i \). Thus, \( \Omega_i\) is the
angular velocity of the black $p$-brane with respect to the angle
\( \phi_i \). Moreover, in the new coordinate system the
off-diagonal metric component
\begin{equation}
\label{newgtphi}
g_{\tilde{t} \tilde{\phi}_i}
= g_{t\phi_i} + \sum_{j=1}^n \Omega_j g_{\phi_j \phi_i }
\end{equation}
has the property that it vanishes at the horizon
\begin{equation}
\label{extrakill}
g_{\tilde{t} \tilde{\phi}_i}\Big|_{r=r_H} = 0
\end{equation}
For completeness we also mention that the new coordinate system
has the Killing vector
\begin{equation}
\label{nullkill}
 V \equiv \frac{\partial}{\partial \tilde{t}} =
\frac{\partial}{\partial t} + \sum_{i=1}^n \Omega_i
\frac{\partial}{\partial \phi_i}
\end{equation}
with norm \( V^2 = g_{\tilde{t}\tilde{t}} \). It then follows from
\eqref{eqkill2} that at the horizon this is a null Killing vector.

The new frame (\ref{newcoor}) also enables us to compute the
temperature. By construction,   both the metric components \( g_{\tilde{t}
\tilde{t}} \) and \( g^{rr} \) are zero for \( r=r_H \). As a
consequence,  the standard procedure of transforming to Euclidean
space can be employed to find the temperature of the $p$-brane:
First we go to the Euclidean signature by a Wick rotation \( \tau
= i\tilde{t} \), and reinterpret the path-integral partition
function as a partition function for a statistical system in $D-1$
dimensions with the temperature $T = 1/ \beta$, where $\beta$ is
the periodicity of $\tau$. This periodicity is determined by
avoiding a singularity in space time (see for example \cite{Hawking:1983dh}).
In the case at hand, the Euclidean metric near the horizon can be
written as
\begin{equation}
ds^2 = - \partial_r g_{\tilde{t}\tilde{t}}|_{r=r_H} ( r-r_H)d\tau^2
+ \frac{1}{\partial_r g^{rr}|_{r=r_H} ( r-r_H)} dr^2 + \cdots
= \rho^2 d\Theta^2 + d\rho^2 + \cdots
\end{equation}
with
\begin{equation}
\rho = 2 \sqrt{ \frac{r-r_H}{\partial_r g_{rr}|_{r=r_H}}},\ \
\Theta = \frac{1}{2} \sqrt{-\partial_r g_{\tilde{t}\tilde{t}}|_{r=r_H}
\partial_r g^{rr}|_{r=r_H}} \tau
\end{equation}
To avoid a conical singularity we need to require that \( \Theta
\) is periodic with period \( 2\pi \), which determines
\begin{equation}
\label{temp} \beta = \frac{1}{T} =  4\pi
\frac{1}{\sqrt{-\partial_r g_{\tilde{t}\tilde{t}}|_{r=r_H}
\partial_r g^{rr}|_{r=r_H}}}
\end{equation}
Using \eqref{extrakill} and \eqref{nullkill} the formula for $T$
can also be written as
\begin{equation}
T = \frac{1}{4\pi} \lim_{r \rightarrow r_H} \sqrt{
\frac{g^{\mu \nu} \partial_\mu V \partial_\nu V}{-V^2} }
\end{equation}
One can then proceed to calculate the temperature $T$ from
(\ref{temp}) using the particular choice of angles \( \theta =
\frac{\pi}{2}\), so that \( \mu_1 = 1 \) and \( \mu_{i \neq 1}  =
0 \). With this choice, we also have $ \partial_\theta \mu_i^2
=
\partial_{\psi_j} \mu_i^2 = 0 $
for all $i$ and $j$. After a tedious calculation, one
obtains
\begin{equation} T = \frac{d-2-2\kappa  }{4\pi r_H \cosh \alpha} \end{equation}
where we have defined
\begin{equation}
\label{kappa} \kappa = \sum_{i=1}^n \frac{l_i^2}{l_i^2 + r_H^2 }
\end{equation}
In fact, it follows\footnote{To see this, define the function $h(x) = 
x^{d-2} \prod_{i=1}^n ( 1 + (l_i/x)^2)  - r_0^{d-2} $ and compute
$h'(x)=  [d-2 - 2 \sum_{i=1}^n l_i^2/(l_i^2 + x^2)] x^{d-3} 
\prod_{i=1}^n ( 1 + (l_i/x)^2) $. 
Using the fact that $h$ and $h'$ are both positive for large $x$, it follows
that $r_H$ cannot be the highest root of $h(x)=0$ if
$d-2- 2 \kappa <0 $.}  
 from \eqref{horizon} that $d-2-2 \kappa \geq 0$ and hence 
$T \geq 0$. 
For $r_0 > 0$ it is thus possible to have $T=0$ and this in turn defines
a boundary on the region of possible values of $(l_1,\ldots,l_n)$ in units
of $r_0$.  

Besides $\Omega_i$ and $T$, the chemical potential
\begin{equation}
\mu = - A_{y^1 y^2 \cdots y^p \tilde{t}} \Big|_{r=r_H} = \tanh \alpha
\end{equation}
is also determined by the solution at the horizon.
The ADM mass $M$ and the charge $Q$ of a spinning black $p$-brane
are the same as for the non-rotating black $p$-brane, since we can
measure these physical quantities in the asymptotic region of the
space-time. In the asymptotic region, one can check that the
metric (\ref{solmet}) does not contain the angular momenta \( l_i \) to
leading order in \( 1/r \). To calculate the ADM mass $M$ one can
therefore use the prescription given in Ref. \cite{Lu:1993vt}.
The angular momenta $J_i$, $i=1 \ldots n$, can be read from the
asymptotic expansion of (\ref{solmet}) using the formula
\cite{Myers:1986un}
\begin{equation}
g_{t\phi_i} = - \frac{8 \pi G}{V_p V(S^{d-1})} \frac{\mu_i^2}{r^{d-2}} J_i
+ \Ord \Big(\frac{\mu_i^2}{r^d} \Big)
\end{equation}
where
\begin{equation}
\label{volsphe}
V(S^{d-1}) = \frac{2 \pi^{d/2}}{\Gamma( d/2 )}
\end{equation}
is the volume of the $d-1$ dimensional unit sphere.
Finally, the entropy $S$ can be calculated from the Bekenstein-Hawking
formula
\begin{equation}
S = \frac{A_H}{4 G }
\end{equation}
where $A_H$ is the area of the outer horizon. Alternatively, $S$
can be found using the integrated Smarr formula reviewed below
which follows from the 1st law of black-hole thermodynamics.

Summarizing, we list the complete set of thermodynamic quantities
for a general spinning black $p$-brane
\begin{subequations}
\label{pbranethermo}
\begin{equation}
\label{admmass}
M = \frac{V_p V(S^{d-1})}{16\pi G} r_0^{d-2}
\Big( d-1 + (d-2)\sinh^2 \alpha \Big)
\end{equation}
\begin{equation}
\label{tands}
T = \frac{d-2-2\kappa }{4 \pi r_H \cosh \alpha} \sp
S = \frac{V_p V(S^{d-1}) }{4G} r_0^{d-2} r_H \cosh \alpha
\end{equation}
\begin{equation}
\label{muandq}
\mu = \tanh \alpha \sp
Q = \frac{V_p V(S^{d-1})}{16\pi G} r_0^{d-2} (d-2) \sinh \alpha \cosh \alpha
\end{equation}
\begin{equation}
\label{angmom}
\Omega_i = \frac{l_i}{(l_i^2 + r_H^2)\cosh \alpha} \sp
J_i = \frac{V_p V(S^{d-1})}{8\pi G} r_0^{d-2} l_i \cosh \alpha
\end{equation}
\end{subequations}
where $\kappa$ is defined in \eqref{kappa}, $V_p$ is the worldvolume of
the $p$-brane and $V(S^{d-1})$ is the volume of the (unit radius)
transverse $(d-1)$-sphere given in \eqref{volsphe}.

In further detail, the internal energy of a spinning charged black
$p$-brane is the mass $M$. The other extensive thermodynamic
parameters are the entropy $S$, the charge $Q$ and the angular
momenta $\{J_i \}$, and the first law of thermodynamics is
\begin{equation}
\label{firstlaw} dM = T dS + \mu dQ + \sum_{i=1}^n \Omega_i dJ_i
\sp  M=M(S,Q,\{J_i \})
\end{equation}
Under the canonical scaling
\begin{equation}
\label{scaling}
r_0 \rightarrow \lambda r_0 \sp
l_i \rightarrow \lambda l_i \sp
\alpha \rightarrow \alpha
\end{equation}
we have the transformations
\begin{equation}
M \rightarrow \lambda^{d-2} M,\ \
S \rightarrow \lambda^{d-1} S,\ \
Q \rightarrow \lambda^{d-2} Q,\ \
J_i \rightarrow \lambda^{d-1} J_i
\end{equation}
It then follows from  Euler's theorem that
\begin{equation}
\label{smarr}
(d-2) M = (d-1) TS + (d-2) \mu Q + (d-1) \sum_{i=1}^n \Omega_i J_i
\end{equation}
which is known as the integrated Smarr formula \cite{Smarr:1973}.
One can also reverse the logic and derive this formula  using
Killing vectors \cite{Myers:1986un}, and then use the scaling
(\ref{scaling}) to find \eqref{firstlaw}. As an important check we
note that the quantities listed in \eqref{pbranethermo} indeed
satisfy \eqref{smarr}.

As an aid to the reader and for use below, we also give here the
explicit expressions for the relevant parameters entering
the solution \eqref{solmet} and the corresponding thermodynamics
\eqref{pbranethermo} for the M-branes in $D=11$ and the D-branes
in $D=10$. To this end, it is useful to define the parameter \( h
\) via the relation
\begin{equation}
\label{hdef}
h^{d-2} = r_0^{d-2} \cosh \alpha \sinh \alpha
\end{equation}
Then, for the branes of M-theory  we have the relations
\begin{equation}
\label{Mrel}
16 \pi G = (2\pi)^8 l_p^9
\sp
h^6 = 2^5 \pi^2 N l_p^6 \;\; \; (\mbox{M2})
\sp
h^3 =  \pi N l_p^3 \;\;\; (\mbox{M5})
\end{equation}
where $l_p$ is the 11-dimensional Planck length and $N$ the number
of coincident branes. In parallel, for the D$p$-branes of type II
string theory we record
\begin{equation}
\label{IIrel}
16 \pi G = (2\pi)^7 g_s^2 l_s^8 \sp h^{d-2} =
\frac{(2\pi)^{d-2} N g_s l_s^{d-2}}{(d-2) V(S^{d-1})} \;\; \;
(\mbox{D}p)
\end{equation}
where $l_s$ is the string length and $g_s$ the string coupling.

\section{Near-horizon limit of general spinning branes
\label{secnhlimit} }

\newcommand{\old}[1]{#1_{\mathrm{old}}}
\newcommand{\oldp}[1]{(#1)_{\mathrm{old}}}

In this section we examine the near-horizon limit of the spinning
$p$-brane solutions considered in Section \ref{secrotbra}. This
provides further insights into the thermodynamics of black branes.
More importantly, this is relevant since according to the
correspondence
\cite{Maldacena:1997re,Itzhaki:1998dd}
between near-horizon brane solutions and field theories, this
gives information about the strongly coupled regime of these field
theories in the presence of non-zero voltages under the
R-symmetry.

In Section \ref{subsecnhsol} we take the near-horizon limit while
in Section \ref{nhtherm} we find the relevant thermodynamic
quantities of the solution. Finally, in Section \ref{subsecdual}
we discuss the map between the supergravity solutions and the dual
field theories, obtaining in particular the free energies of these
field theories.

\subsection{The near-horizon solution
\label{subsecnhsol} }

To find the near-horizon solution, one has to take an appropriate
limit of the solution that is specified as follows: We introduce a
dimensionfull parameter $\ell$ and perform the rescaling
\begin{subequations}
\begin{eqnarray}
\label{rescal}
&& r  = \frac{\old{r}}{\ell^2}   \sp
r_0  = \frac{\oldp{r_0}}{\ell^2}  \sp
l_i = \frac{\oldp{l_i}}{\ell^2} \sp
h^{d-2} = \frac{\old{h}^{d-2}}{\ell^{2d-8}} \\
\label{solrescal}
 && ds^2 = \frac{\oldp{ds^2}}{\ell^{4(d-2)/(D-2)}}
\sp
e^{\phi} = \ell^{2a} e^{\phi_{\rm old}} \sp
A = \frac{A_{\rm old}}{\ell^4} \sp
   G  = \frac{\old{G}}{\ell^{2(d-2)}}
\end{eqnarray}
\end{subequations} %
where  the new quantities on the left hand side are expressed in
terms of the old quantities labelled with a subscript ``old'', and
we recall that $\old{h}$ is defined in \eqref{hdef}. Note that  the
rescaling in \eqref{solrescal} leaves the action \eqref{pbact}
invariant due to the relation \eqref{aeq}. The near-horizon limit
is defined as the limit  \( \ell \rightarrow 0 \) keeping all the
new quantities in (\ref{rescal}) fixed. In
particular, \eqref{rescal} implies that in this limit we have
 $\frac{1}{4} e^{2\alpha} \rightarrow  \ell^{-4}  (h/r_0)^{d-2}$.  

Using \eqref{solmet}-\eqref{solpot} the corresponding near-horizon
solution then becomes
\begin{subequations}
\label{nhsol}
\begin{eqnarray}
\label{nearmet}
ds^2 &=& H^{-\frac{d-2}{D-2}}
\Big( - f dt^2 + \sum_{i=1}^p (dy^i)^2 \Big)
+ H^{\frac{p+1}{D-2}} \Big( \bar{f}^{-1} K_d dr^2
+ \Lambda_{\alpha \beta} d\eta^\alpha d\eta^\beta \Big)
\nn \\ &&
-2 H^{-\frac{d-2}{D-2}}
\frac{1}{K_d L_d} \frac{h^{\frac{d-2}{2}}r_0^{\frac{d-2}{2}}}{r^{d-2}}
\sum_{i=1}^n l_i \mu_i^2 dt d\phi_i
\end{eqnarray}
\begin{equation}
\label{neardil}
e^\phi = H^{\frac{a}{2}}
\end{equation}
\begin{equation}
\label{nearpot} A_{p+1}  = (-1)^p \Big( H^{-1} dt +
\frac{r_0^{\frac{d-2}{2}}}{h^{\frac{d-2}{2}}} \sum_{i=1}^n l_i
\mu_i^2 d\phi_i \Big) \wedge dy^1  \wedge \cdots \wedge dy^p
\end{equation}
\end{subequations}
where the harmonic function is now
\begin{equation}
H = \frac{1}{K_d L_d} \frac{h^{d-2}}{r^{d-2}}
\end{equation}
and the functions $L_d$, $K_d$, $f$, $\bar f$ are as defined before
in \eqref{defs}, \eqref{flattr}, since the scale factor drops out in
these expressions.

\subsection{Thermodynamics in the near-horizon limit
\label{nhtherm} }

We now turn to the thermodynamics of the near-horizon spinning
$p$-brane solution \eqref{nhsol}  obtained in the previous
subsection. Using the rescaling (\ref{rescal})  in the expressions
(\ref{tands}) and (\ref{angmom}) for ($T,S$) and ($\Omega_i,J_i$)
one finds in the near-horizon limit $\ell \rightarrow 0$ the
following quantities,
\begin{subequations}
\label{limthermo}
\begin{equation}
\label{tandsnhlim}
T = \frac{d-2-2\kappa}{4 \pi r_H}
\frac{r_0^{\frac{d-2}{2}}}{h^{\frac{d-2}{2}}} \sp
S = \frac{V_p V(S^{d-1}) }{4G} r_0^{\frac{d-2}{2}} h^{\frac{d-2}{2}} r_H
\end{equation}
\begin{equation}
\Omega_i = \frac{l_i}{(l_i^2 +r_H^2)}
\frac{r_0^{\frac{d-2}{2}}}{h^{\frac{d-2}{2}}} \sp
J_i = \frac{V_p V(S^{d-1})}{8\pi G} r_0^{\frac{d-2}{2}} h^{\frac{d-2}{2}}
l_i
\end{equation}
\end{subequations}
From \eqref{muandq} we see that the chemical potential \( \mu = 1 \)
and that the charge \( Q \) is constant.
Thus, in the near-horizon limit the chemical potential and the
charge are not anymore thermodynamic parameters.

In Appendix \ref{appenergy} we derive the internal energy of a
black $p$-brane in the near-horizon limit by defining this energy
to be the energy above extremality $E = M - Q $. Since $M$ and $Q$
are not affected by the rotation of the brane, it follows from
\eqref{eandfb1} that
\begin{equation}
\label{nhenergy}
E = \frac{V_p V(S^{d-1})}{16\pi G} \frac{d}{2} r_0^{d-2}
\end{equation}

The first law of thermodynamics for a spinning $p$-brane in the
near-horizon limit is
\begin{equation}
\label{nearfl}
dE = T dS + \sum_{i=1}^n \Omega_i dJ_i
\sp
E = E(S,\{ J_i \} )
\end{equation}
Under the canonical rescaling
\begin{equation}
h \rightarrow h,\ \
r_0 \rightarrow \lambda r_0,\ \
l_i \rightarrow \lambda l_i
\end{equation}
we have the transformation properties
\begin{equation}
\label{limscale}
E \rightarrow \lambda^{d-2} E,\ \
S \rightarrow \lambda^{d/2} S,\ \
J_i \rightarrow \lambda^{d/2} J_i
\end{equation}
as follows from \eqref{nhenergy} and \eqref{limthermo}. The
scalings \eqref{limscale} imply with Euler's theorem the
integrated Smarr formula for the near-horizon solution
\begin{equation}
\label{smarrnh}
(d-2) E = \frac{d}{2} TS + \frac{d}{2} \sum_{i=1}^n \Omega_i J_i
\end{equation}
Remark that this conservation law for the near-horizon solution is
different from the Smarr formula \eqref{smarr} of the
asymptotically-flat solution, due to the different scaling
behavior. It is not difficult to obtain the energy function of the
microcanonical ensemble in terms of the extensive variables using
the horizon equation \eqref{horizon} and \eqref{limthermo}, yielding
\begin{equation}
E^{d/2}  = \left(\frac{d}{2} \right)^{d/2} \left( \frac{V_p
V(S^{d-1})}{16\pi G} \right)^{-(d-4)/2} h^{-(d-2)^2/2} \left(
\frac{S}{4 \pi} \right)^{d-2} \prod_i \left( 1 + \left( \frac{2
\pi J_i}{S} \right)^2 \right)
\end{equation}

For later use we also calculate the Gibbs free energy
\begin{equation}
\label{nhgibbs}
F = E - TS - \sum_{i=1}^n \Omega_i J_i = - \frac{d-4}{d} E
= - \frac{V_p V(S^{d-1})}{16\pi G} \frac{d-4}{2} r_0^{d-2}
\end{equation}
which satisfies the thermodynamic relation
\begin{equation}
\label{dfeq}
dF = - S dT - \sum_{i=1}^n J_i d\Omega_i \sp
F= F(T,\{ \Omega_i \} )
\end{equation}
A remark is in order here for the special case $d=4$, which
includes the D5 and NS5-brane in 10 dimensions, since in that case
it follows from \eqref{nhgibbs} that $F=0$. From \eqref{dfeq} we
observe that since the partial derivatives of $F$ with respect to
$T$ and $\{ \Omega_i \}$ are nonzero, these variables cannot be
independent.  Thus the phase diagram in terms of these variables
degenerates into a submanifold with at least one dimension less.
This point will be further illustrated in Section
\ref{subsecgrandcan} where we discuss the phase diagram for one
non-zero angular momentum. For the non-rotating case, one
immediately deduces from \eqref{tandsnhlim}
that the temperature must be constant for $d=4$.

Since the Gibbs free energy is properly given in terms of the
intensive quantities $T$ and $\{\Omega_i \}$ we need to write the
expression \eqref{nhgibbs} in terms of these variables\footnote{In
Section \ref{subsecgrandcan} we obtain the exact expressions
for one non-zero angular momentum.}.
The change of variables from the supergravity
variables $(r_0,\{l_i\})$ to these thermodynamic variables is
given in Appendix \ref{basisch} in a  low angular momentum
expansion
\begin{equation}
\frac{l_i}{r_0} \ll 1
\end{equation}
through order ${\Ord} (l_i^4)$.
Using the result \eqref{r0exp} we find that
\begin{eqnarray}
\label{freeexp} F &=&  - \frac{V_p V(S^{d-1})}{16\pi G}
\frac{d-4}{2} \tilde{T}^{(2d-4)/(d-4)} h^{(d-2)^2/(d-4)} \left[ 1
+ \frac{2}{d-4} \sum_i \tilde \omega_i^2  \right. \nn \\ && \left.
- \frac{2(d-6)}{(d-2)(d-4)^2} \left( \sum_{i} \tilde \omega_i^2
\right)^2 + \frac{1}{d-4} \sum_i \tilde \omega_i^4
  +   \ldots \right]
\end{eqnarray}
for $d \neq 4$ where we have defined $ \tilde{T} = 4\pi T/(d-2)$
and $\tilde \omega_i =\Omega_i/\tilde{T}$.

 As will be explained in Section \ref{subsecdual}, $F$ is the free
energy for the field theory living on the brane in the strongly
coupled large $N$ limit. In Section \ref{secfieldtheory} we
compare this expression with the corresponding expressions in the
weakly coupled field theory. Moreover, in Section \ref{secfreeenergy} 
we show that the free energy \eqref{nhgibbs} is reproduced 
by calculating the (regularized) Euclidean action of the solution.

\subsection{The dual field theories
\label{subsecdual} }

In the remainder of this paper, we will restrict ourselves to the
spinning brane solutions of type II string theory and M-theory.
For these $p$-branes we can map
\cite{Maldacena:1997re,Itzhaki:1998dd}
the near-horizon limit  to a dual Quantum Field Theory (QFT) with
16 supercharges, namely the field theory that lives on the
particular $p$-brane in the low-energy limit. As explained in
Refs.
\cite{Maldacena:1997re,Itzhaki:1998dd},
in the near-horizon limit the bulk dynamics
decouples\footnote{With the exception of the D6-brane, as
discussed for example in \cite{Itzhaki:1998dd}.} from the field
theory living on the $p$-brane, so that the supergravity solution
in the near-horizon limit describes the strongly coupled large $N$
limit of the dual QFT.

The fact that the branes are spinning, introduces the new
thermodynamic parameters \( \Omega_i \) and \( J_i \) on the
supergravity side which need to be mapped to the field theory
side, where they are  conjectured to correspond to voltage and
charge for the field theory R-symmetry group. Thus, the validity
of this correspondence requires the R-symmetry groups to be
$SO(d)$, with the charges $J_i$ taking values in the Cartan
subgroup $SO(2)^n$ of $SO(d)$ and their Legendre transforms
corresponding to the voltages $\Omega_i$. In Section
\ref{secfieldtheory} we analyze the field theory in the weakly
coupled regime using the R-charge quantum numbers of the massless
degrees of freedom.

Indeed, for the $p$-branes with $p \leq 6$  it has been noted in
\cite{Itzhaki:1998dd,Boonstra:1998mp} that the dual QFTs have the
correct R-symmetry groups. In particular, in Ref.
\cite{Itzhaki:1998dd} it was noted that D$p$-branes
 have an \( ISO(1,p) \times SO(d) \) symmetry in the
 near-horizon limit, where  \( SO(d) \) corresponds to the R-symmetry
group of the dual field theory and  \( ISO(1,p) \) corresponds
to the Poincar\'e symmetry of the dual field theory. Moreover, in
the dual frame, as considered in Ref. \cite{Boonstra:1998mp} (see
also  \cite{Behrndt:1999mk}) the near-horizon solutions under
consideration can be written as \( DW_{p+2} \times S^{d-1} \) with
a linear dilaton field, where \( DW_{p+2} \) is the $p+2$
dimensional Domain-Wall. Also in this language, the isometry group
\( SO(d) \) of \( S^{d-1} \) translates into the R-symmetry group
of  the dual field theory.

As explained in \cite{Itzhaki:1998dd,Behrndt:1999mk} we can trust
the supergravity description of the dual field theory, when the
string coupling \( g_s \ll 1 \) and the curvatures of the geometry
are small. This implies in all cases that the number of coincident
$p$-branes $N \gg 1$. For the M2- and M5-brane this is the only
requirement   since there is no string coupling in 11-dimensional
M-theory. For the D$p$-branes in 10 dimensions one must further
demand that \cite{Itzhaki:1998dd}
\begin{equation}
\label{geff}
 1 \ll g_{\mathrm{eff}}^2 \ll N^{\frac{4}{7-p}} \sp
 g_{\mathrm{eff}}^2 = g_{\mathrm{YM}}^2 N r^{p-3}
\end{equation}
where $ g_{\mathrm{eff}}^2 $
is the effective coupling, \( g_{\mathrm{YM}}^2 \) the coupling of
the Yang-Mills theory on the D$p$-brane and \( r \) is the
rescaled radial coordinate in the near-horizon limit (the distance
to the D-brane probe) and the Higgs expectation value in the dual
QFT\footnote{See e.g.
\cite{Itzhaki:1998dd,Tseytlin:1998cq,Kiritsis:1999ke,Kiritsis:1999tx,Cai:1999ad}
for discussions of D-brane probes, including  thermal and spinning
D-branes.}. Thus, the near-horizon limit describes the dual QFT in
the large $N$ and strongly coupled limit. Note that the thermodynamic
expressions are valid when $r$ is replaced by $r_H$ in Eq. \eqref{geff}.  
For larger values of the
effective coupling the D1 and D5-brane flow to the NS1 and NS5-brane
respectively, while
the self-dual D3-brane flows to itself \cite{Itzhaki:1998dd}.
In particular, the NS1-brane description is valid for
$ N^{2/3} \ll  g_{\mathrm{eff}}^2 \ll N $ and the NS5-brane description
for $ N^{2} \ll  g_{\mathrm{eff}}^2 $ (see also
Refs.
\cite{Maldacena:1997cg,Sfetsos:1999pq} for further details on
the type II NS5-branes).

In view of this correspondence, we can  write the Gibbs free
energy and other thermodynamic quantities in terms of field theory
variables\footnote{See \cite{Behrndt:1999mk} for a more detailed
explanation of the mapping between the near-horizon supergravity
solutions and QFTs, including a description of the cases with $D <
10$.}. In particular we need to specify the relation between the
parameter $\ell$ entering the near-horizon limit and the relevant
length scale of the theory and compute the rescaled quantities in
\eqref{rescal}. In the following $N$ is the number of coincident
branes and we have defined the quantity $\omega_i =\Omega_i/T$.

For the M2-brane we need the relations
\begin{equation}
\label{M2rel}
\ell = l_p^{3/4} \co 16 \pi G = (2\pi)^8 \sp h^6 = 2^5 \pi^2 N
\end{equation}
where \( l_p \) is the 11-dimensional Planck length and we have
used \eqref{Mrel}. Using this in \eqref{freeexp} gives the Gibbs free
energy
\begin{equation}
\label{strongm2fe} F_{\rm M2} = -\frac{2^{7/2} \pi^2}{3^4} N^{3/2}
V_2 T^3 \left[ 1 +  \frac{9}{8 \pi^2} \sum_{i=1}^4 \omega_i^2
 - \frac{27}{128 \pi^4 }
\left( \sum_{i=1}^4 \omega_i^2 \right)^2
 + \frac{81}{64 \pi^4 }
\sum_{i=1}^4 \omega_i^4  + \ldots \right]
\end{equation}
For the M5-brane we have
\begin{equation}
\label{M5rel}
\ell = l_p^{3/2} \co  16 \pi G = (2\pi)^8 \sp h^3 = \pi N
\end{equation}
giving
\begin{equation}
\label{strongm5fe} F_{\rm M5} = -\frac{2^6 \pi^3}{3^7} N^3  V_5
T^6 \left[ 1 +  \frac{9}{8\pi^2 }  \sum_{i=1}^2 \omega_i^2
 + \frac{27}{128\pi^4   }
\left( \sum_{i=1}^2 \omega_i^2 \right)^2
 + \frac{81}{256\pi^4   }
\sum_{i=1}^2\omega_i^4  + \ldots \right]
\end{equation}
For the D$p$-brane of  type II string theory
 we have from \eqref{IIrel}
\begin{equation}
\label{dprel}
\ell = l_s \co h^{d-2} = \frac{(2\pi)^{2d-9} }{(d-2)V(S^{d-1})}
\lambda \sp \frac{V(S^{d-1}) h^{2(d-2)} }{16\pi G} =
\frac{(2\pi)^{2d-11}}{(d-2)^2 V(S^{d-1}) } N^2
\end{equation}
where \( l_s \), $g_s$ are the string length and coupling and
$\lambda= g_{\mathrm{YM}}^2 N$ is the 't Hooft coupling with the
 Yang-Mills coupling given by \(
g_{\mathrm{YM}}^2 = (2\pi)^{p-2} g_s l_s^{p-3} \). Using these
relations in \eqref{freeexp} we obtain  (for \( p \neq 5 \))
\begin{equation}
\label{strongdpfe} F_{{\rm D}p} = - c_p V_p N^2
\lambda^{-\frac{p-3}{p-5}} T^{\frac{2(7-p)}{5-p}} \left[ 1 +
\frac{S^1_p}{\pi^2 }
\sum_i \omega_i^2
+ \frac{S^2_p}{\pi^4 }
 \left( \sum_{i} \omega_i^2 \right)^2
+ \frac{S^3_p}{\pi^4 } \sum_i
\omega_i^4
  +   \ldots \right]
\end{equation}
where $c_p$, $S^1_p$, $S^2_p$ and $S^3_p$ are listed in
Table \ref{tabfrdata}
and we recall that $p = 9-d$ for $D=10$.

Under type IIB S-duality we have
$\tilde g_s = 1/g_s$ and $\tilde l_s = l_s g_s^{1/2}$, so that for
the type IIB NS1 and NS5-brane we need  $\ell = \tilde l_s$
and the thermodynamics is exactly the same as for the D1 and D5-brane
when expressed in terms of $\lambda$ and $N$.
Note that the free energies
for the D$p$-branes with $p \leq 4$ are negative, while the D5 and NS5-brane
have zero free energy and the D6-brane positive free energy.

In the discussion above, we have chosen to explicitly write down the free
energies in terms of the variables of the dual field theories, since these
expressions will play an important role below. Of course, the same can 
be done for the other thermodynamic quantities listed in 
\eqref{limthermo} using
\eqref{M2rel}, \eqref{M5rel} and \eqref{dprel}. Note also that
for the special value of $\kappa = \frac{1}{2} (d-2)$ the temperature
vanishes, implying that besides the usual extremal limit describing
zero temperature field theory, we also have a limit in which the temperature
is zero, accompanied by non-zero R-charges.

\begin{table}
\begin{center}
\begin{tabular}{|c||c|c|c|c|}
\hline
$p$ & $c_p$ & $S^1_p$ & $S^2_p$     &  $S^3_p$  \\ \hline \hline
0   & $(2^{21} 3^2 5^7 7^{-19} \pi^{14})^{1/5} $
    & $\frac{49}{40}$ & $-\frac{1029}{3200}$ & $\frac{2401}{1280}$  \\ \hline
1   &  $2^4 3^{-4} \pi^{5/2}  $
    & $\frac{9}{8}$ & $-\frac{27}{128}$ & $\frac{81}{64}$ \\ \hline
2   & $(2^{13} 3^5 5^{13}  \pi^{8})^{1/3} $
    & $\frac{25}{24}$ & $-\frac{125}{1152}$ & $\frac{625}{768}$ \\ \hline
3   & $2^{-3} \pi^2 $
    & $1 $ & $0$ & $\frac{1}{2}$ \\ \hline
4   &  $2^5 3^{-7} \pi^{2}  $
    & $\frac{9}{8}$ & $\frac{27}{128}$ & $\frac{81}{256}$ \\ \hline
6   &  $-2^3 \pi^4$
    & $-\frac{1}{8}$ & $0 $ & $\frac{5}{256}$ \\ \hline \hline
\end{tabular}
\end{center}
\caption{Relevant coefficients for the free energy of D$p$-branes.
\label{tabfrdata} }
\end{table}

\section{Stability analysis of near-horizon spinning branes
\label{seccritbeh} }

\newcommand{\hes}{\mbox{Hes}}

In this section we analyze the critical behaviour of the
near-horizon limit of spinning $p$-branes, using the
thermodynamics obtained in Section \ref{nhtherm}. Using the
mapping between the supergravity solutions and the dual QFTs,
as described in Section \ref{subsecdual}, we can
find the critical behaviour for the strongly coupled dual field
theories with non-zero voltages under the R-symmetry.

Section \ref{subsecbound} presents a general discussion of
boundaries of stability in the grand canonical and canonical
ensemble. These two ensembles are then considered in more detail
in Sections \ref{subsecgrandcan} and \ref{subseccan} respectively.
In Section \ref{subseccritexp}
 we finally consider the critical exponents in
the two ensembles.

\subsection{Boundaries of stability
\label{subsecbound} }

There are two different settings in which we can study the
stability of near-horizon spinning branes. In the first one, to
which we refer as the grand canonical ensemble, we imagine the
system to be in equilibrium with a reservoir of temperature $T$
and angular velocities $\Omega_i$. Thermodynamic stability then requires
negativity of the eigenvalues of the Hessian of the Gibbs free
energy. In particular, a boundary of stability occurs when the
determinant of the Hessian is zero or infinite. In the second one,
referred to below as the canonical ensemble, we have constant
angular momenta $J_i$ and a heat reservoir with temperature $T$.
Stability in this situation demands positivity of the heat
capacity $C_J$ and the boundaries of stability occur when this
specific heat is zero or infinite.

In the case at hand, we have a system in which the thermodynamic
quantities are given in terms of the supergravity variables, so
that the boundaries of stability will crucially depend on the
change of variables between these two descriptions. We will
therefore repeatedly need the determinants of the Jacobians, and
we define \( D_{T \Omega} \) as the determinant \( \frac{
\partial(T,\Omega_1,\Omega_2,...,\Omega_n)}
{\partial(r_H,l_1,l_2,...,l_n)} \) and likewise for \( D_{T J} \),
\( D_{S \Omega} \) and \( D_{S J} \).

 In the grand canonical ensemble
we need the determinant of the Hessian of the Gibbs free energy,
which can be written as\footnote{One could also
use the determinant of the Hessian of the internal energy
$E(S,J)$, which is the inverse of the Hessian of the Gibbs free
energy.}
\begin{equation}
\det \hes (-F) = \frac{D_{S J }} {D_{T\Omega}}
\end{equation}
so that the zeroes of the two determinants $D_{S J}$,
$D_{T\Omega}$ determine the boundaries of stability. For
completeness and use below we also give the specific heat
\begin{equation}
C_\Omega = T \Big( \frac{\partial S}{\partial T}
\Big)_{\Omega_1,...,\Omega_n} = T\frac{D_{S \Omega }}
{D_{T\Omega}}
\end{equation}
showing that $\det \hes(F)$ and $C_\Omega$ may have different
zeroes. In the canonical ensemble, on the other hand, we need the
specific heat
\begin{equation}
 C_J = T \Big( \frac{\partial S}{\partial T} \Big)_{J_1,...,J_n}
 = T \frac{D_{S J}}{D_{T J}}
\end{equation}
and the boundaries of stability are determined by the determinants
$D_{SJ}$ and $D_{TJ}$. 

In further detail, using the thermodynamic quantities in
\eqref{limthermo} it then follows that
\begin{subequations}
\begin{equation}
\det \hes (-F) = 8 \pi^2
\left( \frac{V_p V(S^{d-1}) h^{d-2} }{8 \pi G}  \right)^{n+1}
 r_H^4 \prod_i (1+x_i)^2  \frac{\Delta_{S J} }{\Delta_{T\Omega}}
\end{equation}
\begin{equation} \label{com} C_\Omega = \frac{V_p
V(S^{d-1})}{4G} h^{\frac{d-2}{2}} r_0^{\frac{d-2}{2}} r_H
(d-2-2\kappa)  \frac{\Delta_{S\Omega}} {\Delta_{T\Omega}}
\end{equation}
\begin{equation}
\label{cj}
C_J
= \frac{V_p V(S^{d-1})}{4G} h^{\frac{d-2}{2}} r_0^{\frac{d-2}{2}} r_H
(d-2-2\kappa ) \frac{\Delta_{S J}}{\Delta_{T J}}
\end{equation}
\end{subequations}
where the functions $\Delta$ are related to the determinants $D$ up
to positive define functions, and given by
\begin{subequations}
\label{Dfun}
\begin{eqnarray}
\label{dto}
\Delta_{T \Omega} &=& (d-4) \Big[ d-2 - (d-4)  \sum_i
x_i + (d-6) \sum_{i < j} x_i x_j  \nn  \\ &&  -(d-8) \sum_{i < j <
k} x_i  x_j x_k
 +(d-10) x_1 x_2 x_3 x_4 \Big]
\end{eqnarray}
\begin{eqnarray}
\label{dso}
\Delta_{S \Omega} &=& d - (d-4)  \sum_i x_i +
(d-8)\sum_{i
< j} x_i x_j \nn \\ && -(d-12)\sum_{i < j < k} x_i  x_j x_k
+(d-16) x_1 x_2 x_3 x_4
\end{eqnarray}
\begin{eqnarray}
\label{dtj}
\Delta_{T J} &=&
 (d-4)(d-2)
- 2(d-8) \sum_i \frac{x_i}{1+x_i}  \nn \\ && - 4(d-2) \sum_{i}
\frac{x_i^2}{(1+x_i)^2}  + 16 \sum_{i<j}
\frac{x_i}{1+x_i}\frac{x_j}{1+x_j}
\end{eqnarray}
\begin{equation}
\label{dsj}
\Delta_{S J} = d
\end{equation}
\end{subequations}
Here, we have defined the dimensionless ratios
\begin{equation}
\label{xdef}  x_i = \frac{l_i^2}{r_H^2}
\end{equation}
The expressions are written for the case of maximal possible
number of angular momenta $n=4$, but hold also for $n<4$ by
setting the appropriate $x_i=0$. One observes that, as seen for
the free energy in \eqref{nhgibbs}, the case \( d=4 \) is  special
since \( \Delta_{T \Omega} = 0 \) identically, implying that the
coordinates \( (T,\{\Omega_i\}) \) are not independent.

Since $\Delta_{SJ} \neq 0$, the boundaries of stability in the two
ensembles can thus be determined as follows: In the grand
canonical ensemble a boundary is reached when $\Delta_{T
\Omega}=0$ and the Hessian of the Gibbs free energy diverges. In
the canonical ensemble on the other hand, we have a boundary of
stability when  $\Delta_{TJ}=0$, in which case the specific
heat $C_J$ diverges. More precisely, the boundaries of stability
are $n$-dimensional submanifolds in the $(n+1)$-dimensional phase
diagram, where one of these two determinants vanish. In the
following subsections we study these conditions for the special
case of $m \leq n$ equal angular momenta, supplemented with a
detailed discussion for the simplest case of one non-zero angular
momentum. For the non-dilatonic branes this analysis was performed
in Refs. \cite{Gubser:1998jb,Cai:1998ji,Cvetic:1999rb}.

It should be remarked that, at first sight, the analysis shows
that we do not have any first-order phase transitions, since all
first derivatives of the thermodynamic potentials are continuous
everywhere. The phase transitions are instead second-order, though
this result should be taken with care, since  \cite{Cvetic:1999rb}
has given evidence for a first-order phase transition. We will
comment on this possibility in the next subsection. Another
general result of the analysis is that the boundaries of stability
are distinct in the two ensembles that we consider, with
a larger region of stability in the canonical ensemble as compared
to the grand canonical ensemble, in accordance with standard
thermodynamics.

We emphasize here that although we have phrased the analysis in
terms of the variables $(r_H,x_i)$,  these are in one-to-one
correspondence with the thermodynamic extensive variables
$(S,J_i)$ through the relations
\begin{equation}
\label{sjrel} \sqrt{x_i} =  \frac{2\pi J_i}{S} \sp r_H^{d/2} =
\left( \frac{V_p V(S^{d-1})}{16 \pi G} \right)^{-1} h^{-(d-2)^2/2}
\frac{S}{4\pi}\prod_i \sqrt{1 + x_i}
\end{equation}
which follow from \eqref{limthermo} and \eqref{xdef}.
 Hence, conditions on $x_i$ can be
directly translated into conditions on the ratio $J_i/S$.
Alternatively, one may rephrase the stability conditions in terms
of the dimensionless ratios \label{chidef} \begin{equation} \chi_i
= \frac{E^{d/2}}{J_i^{d-2}} = d^{d/2} 2^{4-3d}  \left(\frac{V_p
V(S^{d-1})}{16 \pi G}\right)^{-(d-4)/2} h^{-(d-2)^2/2}
 \left(
\frac{r_0}{l_i} \right)^{d-2} \end{equation} where
\begin{equation}
 \left(
\frac{r_0}{l_i} \right)^{d-2} = x_i^{(d-2)/2} \prod_{j=1}^n
(1+x_j)
\end{equation}
For the case of one angular momentum, $\chi$ is up to a numerical
constant the variable used in the D3-brane analysis of Refs.
\cite{Gubser:1998jb,Cvetic:1999rb}.

\subsection{Grand canonical ensemble \label{subsecgrandcan}  }

We  consider the case of $m \leq n$ equal non-zero angular
momentum, so that \( x_i = x=l^2/r_H^2\), $i=1 \ldots m$, in which
case the relevant quantity $\Delta_{T \Omega}$ in \eqref{dto}
simplifies to
\begin{equation}
\label{com1} \Delta_{T\Omega} = (d-4) \Big(d-2 - (d-2 -2m) x \Big)
(1-x)^{m-1}
\end{equation}
We first note that for $d=3$ (which includes the D6-brane) $\det
\hes (-F)$ is less than zero for all $x$, and  hence corresponds
to an unstable situation. The case $d=4$ (which includes the D5
and NS5-brane), for which $T$ and $\Omega_i$, $i=1 \ldots n$ are
not independent will be treated separately at the end of this
subsection, so in the following we assume $d \geq 5$. In this
case, we know that for zero angular momentum, i.e. $x=0$,  the
branes are stable. We will be concerned only with the first
instability that occurs as $x$ is increased,  which is hence
determined by the first zero of $\Delta_{T \Omega}$.

It follows from \eqref{com1} that there is a boundary of stability
at the value
\begin{equation}
\label{critgce}
 x_c^{(m)} = \left\{ \begin{array}{ll}
 \frac{d-2}{d-4} & \sp m =1 \\
 1 & \sp m > 1 \end{array}\right.
\end{equation}
In further detail, stability requires $x \leq x_c^{(m)}$ or
equivalently, using \eqref{sjrel} this becomes
\begin{equation}
J \leq \sqrt{x_c^{(m)}}
\frac{S}{2\pi} 
\end{equation}
One may also calculate from \eqref{limthermo} that for
$m$ equal angular momenta
\begin{equation} \label{mgentom} \tilde \omega =
\frac{\sqrt{x}}{1+ x/x^{(m)}_\star }
\end{equation}
where we recall the definitions $\tilde T = 4 \pi T/(d-2) $,
$\tilde \omega = \Omega/\tilde T$ and we have defined
\begin{equation}
\label{critvalgce}
 x^{(m)}_\star = \frac{d-2}{d-2-2m}
\end{equation}
If for instance $S(T,\{ \Omega_i\})$ is known,  
Eq.  \eqref{mgentom} can be viewed as an equation of state
using $\sqrt{x}  = 2 \pi J/S$.  
With the
critical values of $x$ in \eqref{critgce} the corresponding
critical values of $\tilde \omega$ are determined by substitution
in  \eqref{mgentom} so that
\begin{equation}
 \tilde \omega_c^{(m)} = \left\{ \begin{array}{ll}
 \frac{1}{2} \sqrt{\frac{d-2}{d-4}} & \sp m =1 \\
 \frac{d-2}{2(d-2-m)} & \sp m > 1 \end{array}\right.
\end{equation}
summarized together with $x_c^{(m)}$ in Table \ref{tabcrit}. As
seen from the table, the critical values $\tilde \omega_c^{(m)}$
increase as the number of non-zero angular momenta increases, so
turning on more
 equal-valued angular momenta has a stabilizing effect.

\und{Specific heat}

It is also interesting to examine the behavior of the specific
heat $C_\Omega $ in \eqref{com}, for which we need in addition to
\eqref{com1}, \begin{subequations}\label{com2}
\begin{equation}
\Delta_{S\Omega} = \Big( d - (d-4 m ) x \Big)(1-x)^{m-1}
\end{equation}
\begin{equation}
 (d-2-2 \kappa) = \frac{1}{(1+x)^m}[d-2 + (d-2 -2 m) x]
\end{equation}
\end{subequations}
which follows from \eqref{dso} and $\kappa$ in \eqref{kappa}.
Besides a diverging specific heat at $x_c^{(m)}$, we see that
$C_\Omega$ vanishes, on the other hand, for the values
\begin{equation}
\label{specheat} x_0^{(m)}=\frac{d}{d-4m}
  \sp  \sp x_T^{(m)} =- x^{(m)}_\star
\end{equation}
obtained from the zero of $\Delta_{S \Omega}/\Delta_{T \Omega}$
and the temperature $T$ using \eqref{com2}\footnote{Note also that
for $d=3$ the specific heat vanishes at $x_T =1$.}.

\und{One non-zero angular momentum}

In the remainder of this subsection 
we restrict to the case of one non-zero angular
momentum, which by itself exhibits various interesting physical
phenomena. The stable region is $x \leq x_c$, with the critical
value given by
\begin{equation}
x_c \equiv x_c^{(1)} = \frac{d-2}{d-4} \end{equation} or using
\eqref{sjrel},
\begin{equation}
 J \leq \sqrt{\frac{d-2}{d-4}} \frac{S}{2\pi}
\end{equation}
The stability requirement thus sets an upper bound on the angular
momentum, and a phase transition occurs at the critical value of
the angular momentum density. In the dual  field theory, this
corresponds to a critical value of the R-charge density.
 From eqs. \eqref{limthermo} one can
also derive the general formulae
 \begin{subequations}
\begin{equation}
\label{gentom}  \tilde \omega = \frac{\sqrt{x}}{1+ x/x_c }
\end{equation} \begin{equation}
\label{trel} \frac{1}{\tilde T} =  \frac{\sqrt{1+x}}{1+ x/x_c }
r_H^{(4-d)/2} h^{(d-2)/2} 
\end{equation}
 \end{subequations}
It is not difficult to see that at the boundary of stability
$x=x_c $ where the Hessian  diverges, the ratio $\tilde \omega$ is
maximized\footnote{As a consequence, one could have determined
this boundary of stability by maximizing $\Omega/T$, providing an
alternative method without having to resort to computing
Jacobians.}, so that the supergravity description sets an upper
bound on this quantity,
\begin{equation}
\label{crittom} \tilde \omega \leq \tilde \omega_c = \frac{1}{2}
\sqrt{x_c}
\end{equation}
Moreover, as easily seen from \eqref{gentom}, for each value of
$\tilde \omega$ below this maximum there are two values of $x$,
one corresponding to a stable  and the other to an unstable
configuration. In particular the two supergravity descriptions
with $(r_H,x)$ and $(\tilde r_H, \tilde x)$ related by
\begin{equation}
\label{sheets}
 \tilde x = \frac{x_c^2}{x} \sp \tilde r_H = r_H \left(
\frac{x^2}{x_c^2} \frac{1 + \frac{x_c^2}{x} }{1 + x }
\right)^{1/(d-4)}
\end{equation}
give the same values of $T$, $\Omega$. The phase diagram therefore
consists of two sheets, a stable one and an unstable one.

As an illustration consider a process in which one starts with a
non-rotating non-extremal brane at given $r_0 = r_H $ and turn on
the angular momentum $l$ adiabatically, while keeping the horizon
radius constant.   When the critical value $\tilde \omega_c$ is
reached the configuration becomes unstable and for the D3-brane
two scenarios have been proposed \cite{Cvetic:1999rb}: D-brane
fragmentation, in which the branes fly apart in the transverse
dimension, and phase mixing in which angular momentum localizes on
the brane. The latter possibility will be briefly discussed below
for the general $p$-brane. Note also that for vanishing horizon
radius but non-zero angular momentum we have that $x \rightarrow
\infty$, so that the brane is unstable in this situation, and
turning on adiabatically the horizon radius would not cure this
instability. Note, however, that $C_\Omega$ is positive not only
for $x
<x_c$ but also for $x > x_0 \equiv x_0^{(1)}$ in \eqref{specheat},
so the specific heat will be positive in this situation.

\begin{table}
\begin{center}
\begin{tabular}{|c|c||c|c||c|c||c||}
\hline & &  \multicolumn{2}{|c||}{GCE}  &
 \multicolumn{2}{c||}{CE} &  \\
 \cline{3-6} $d$ & $m$
& $x_c^{(m)}$ & $\tilde \omega_c^{(m)} $ & $\hat x_c^{(m)}$ &
$\tilde j_c^{(m)}$ & $x_T^{(m)}$
\\ \hline \hline 5 & 1 & 3 & $\frac{\sqrt{3}}{2}$ & $2 + \sqrt{5}$ &
7.238  &
\\ \cline{2-7}    & 2   &  1 & $\frac{3}{2}$ 
&  & & 3
\\ \hline \hline  6 & 1 & 2 &$\frac{\sqrt{2}}{2}$
&  $\frac{5+\sqrt{33}}{2} $ & 14.12  &  \\ \cline{2-7}  & 2 & 1 &
1 &
 & &
\\ \cline{2-7}
  & 3  & 1     & 2       &    & & 2
\\ \hline \hline 7  &  1  &  $\frac{5}{3}$ &  $\frac{\sqrt{15}}{6}$   &
  $\frac{16+\sqrt{301}}{3} $  & 49.59 &
   \\ \cline{2-7}     &  2  &  1 & $\frac{5}{6}$
   & $\frac{17}{5}+\frac{2}{5}\sqrt{91}$   & 165.5  & \\ \cline{2-7}
& 3   & 1 & $\frac{5}{4}$       &     & & 5 \\ \hline \hline 8   &
1 & $\frac{3}{2}$ & $\frac{\sqrt{6}}{4}$ &  & &
\\ \cline{2-7}    &  2  &  1    &    $\frac{3}{4}$     &
 $3+2\sqrt{3}$ & 211.5 &
\\ \cline{2-7}    &  3  & 1 & 1       &    &  & \\
\cline{2-7}    &  4  &  1    &  $\frac{3}{2}$ &   &  & 3 \\ \hline
\hline 9 & 1 & $\frac{7}{5}$ &  $\frac{\sqrt{35}}{10}$     &    &
&
\\ \cline{2-7}
 &  2  & 1 &  $\frac{7}{10}$
 & $\frac{11}{3}+\frac{2}{3}\sqrt{39}$     & 522.2 &
 \\ \cline{2-7}   & 3  &  1    &  $\frac{7}{8}$
 & $\frac{32}{7}+\frac{3}{7}\sqrt{141}$  & 4589 &
\\ \cline{2-7}    & 4   & 1 & $\frac{7}{6}$ &   &  & 7 \\ \hline \hline
\end{tabular}
\end{center}
\caption{Boundaries of stability in the grand canonical ensemble
(GCE) and canonical ensemble (CE) for $m \leq n$ equal non-zero
angular momenta. The values in the last column give zero
temperature. \label{tabcrit} }
\end{table}

\und{Free energy on the two branches}

It is possible to obtain a closed form expression for the free
energy on the two  branches $x \leq x_c$ and $x \geq x_c$
respectively. To this end we solve \eqref{gentom} for $x$ yielding the
two solutions
\begin{equation}
\label{xsol} x_{\pm} = 8  \frac{\tilde \omega_c^4}{\tilde
\omega^2} \left( 1 - \frac{1}{2} \left( \frac{\tilde
\omega}{\tilde \omega_c} \right)^2 \pm \sqrt{ 1- \left(
\frac{\tilde \omega}{\tilde \omega_c} \right)^2 } \right)
\end{equation}
where we have used the value of $\tilde \omega_c$ in
\eqref{crittom}. It is easy to check that the solution $x_-$ has
the property that $x_- \rightarrow 0$ when $\omega \rightarrow 0$,
whereas the other solution $x_+$ goes to infinity in that limit.
Thus, $x_-$ describes the stable branch $0
< x\leq x_c$ and $x_+$ the unstable branch $x > x_c$. To obtain
the explicit expression for the free energy we use \eqref{trel} to
express $r_H$ in terms of $T$ and $x$, as well as \eqref{nhgibbs},
which together with the horizon equation \eqref{horizon} implies
$F \sim (1+x) r_H^{d-2}$. The resulting free energy for each of
the two branches is then
\begin{equation} F_{\pm} = - \frac{V_p V(S^{d-1})}{16\pi G} h^{(d-2)^2/(d-4)}
\frac{d-4}{2} \tilde{T}^{(2d-4)/(d-4)}
(1+x_{\pm})^{2(d-3)/(d-4)} \left( 1 + \frac{x_{\pm} }{x_c}
\right)^{-2\frac{d-2}{d-4}}
\end{equation}
with $x_{\pm}$ given in \eqref{xsol}. As a check, we note that
expanding $F_-$ for small $\tilde \omega$ reproduces the expansion
given in \eqref{freeexp}, as it should. Differentiating $F_-$ with respect 
to $T$ gives the entropy $S(T,\Omega)$, so that we can use $\sqrt{x} 
= 2 \pi J/S$ in \eqref{gentom} to determine the exact form of the equation
of state for one non-zero angular momentum. 
As a curiosity we also
mention the expansion of the free energy $F_+$  on the unstable
branch,
\begin{equation}
F_+ \sim T^2 \Omega^{4/(d-4)} \left[ 1 + {\Ord} \left(\frac{\Omega}{
T} \right) \right]
  \end{equation}
exhibiting a universal $T^2$ dependence for all branes, but due to
the unstable nature of this branch the relevance of this
expression is presently unclear.

\und{Phase Mixing}

In Ref. \cite{Cvetic:1999rb}, it was shown that  for the spinning
D3-brane there exists a possibility that a mixing of these two
phases is thermodynamically favored (maximizing entropy), so that
as a consequence of the instability angular momentum is localized
on the brane. To carry out this analysis  for the general case $d
>4$, one needs to work in the microcanonical ensemble and consider
the mixed states determined by \eqref{sheets}. Thus the problem is
to maximize the entropy
\begin{equation}
S_{\rm av} = \mu  S (r_H,x) + (1 - \mu ) S (\tilde r_H, \tilde x)
\end{equation}
for given energy $E_{\rm av}$ and angular momentum $J_{\rm av}$,
subject to the constraints
\begin{equation}
E_{\rm av}  = \mu  E (r_H,x) + (1 - \mu ) E (\tilde r_H, \tilde x)
\sp
J_{\rm av} =  \mu  J (r_H,x) + (1 - \mu ) J (\tilde r_H, \tilde x)
\end{equation}
with $(\tilde r_H, \tilde x)$ expressed in $(r_H,x)$ through
\eqref{sheets}. We have not carried out this analysis but expect
that the features observed for $d=6$ (including a first-order
phase transition) in \cite{Cvetic:1999rb}, will persist for the
other cases $d
> 4$. We thus expect that there will be a mixed state and
first-order phase transition at some critical value of $x < x_c$.

\und{The case $d=4$}

As pointed out before, the case $d=4$ needs a special treatment,
since from \eqref{nhgibbs} we have that the free energy vanishes,
so that $d F=0$. This is due to the fact that the $n+1$ variables
$(T,\{\Omega_i \})$ are not independent anymore. Indeed, for one
non-zero angular momentum we read off from \eqref{limthermo} that
\begin{equation}
\label{curve} ( 2 \pi T)^2 + \Omega^2 = h^{-2} \end{equation}
which characterizes the phase space. For zero angular momentum we
recover the known fact that the temperature is constant
for the NS5 and D5-branes. As a further check, using also
$\Omega/T = (2 \pi)^2 J/S$ in this case, the curve \eqref{curve}
implies that $S d T + J d \Omega =0$, in accord with the
thermodynamic relation \eqref{dfeq} with $dF=0$.

\subsection{Canonical ensemble \label{subseccan} }

We also discuss the canonical ensemble for $m$ non-zero equal
angular momenta, for which the relevant quantity $\Delta_{TJ}$
takes the form,
\begin{eqnarray}
\label{cj1} \Delta_{TJ} & =  \frac{1}{(1+x)^2}  & \Big[ (d-2)(d-4)
+ 2\Big( (d-2)(d-4) -( d-8)m \Big) x   \nonumber \\
 && + (d-2-2m)(d-4-4m)x^2 \Big]
\end{eqnarray}
and we recall that the specific heat $C_J$ also vanishes at the
zeroes of the temperature, i.e. $x_T^{(m)}$ given in
\eqref{specheat}. The case $d=4$ is stable for any $m$ and for
$d=3$ we find the curious behavior that there is lower bound on
$x$ namely $(-4 +\sqrt{21})/5$, so that e.g. the
non-rotating D6-brane is unstable, but becomes stable
in the canonical ensemble when the
angular momentum is large enough. In the following we will
restrict again to $d \geq 5$.

The positive solutions of the quadratic equation \eqref{cj1} are
listed as $\hat x_c^{(m)}$ in Table \ref{tabcrit}, and correspond
to the boundaries of stability, with the property that for $x\leq
\hat x_c^{(m)}$ the branes are stable. From the table we infer a
number of observations: For one non-zero angular momentum the
branes with $d=8,9$ are stable for any value of $x$, but when more
angular momenta are switched on a boundary of stability emerges.
Moreover, for maximal number of non-zero angular momenta all
branes are stable.

To further examine the boundary of stability we use
\eqref{limthermo} to construct the dimensionless ratio
\begin{subequations}
\begin{eqnarray}
&\tilde j = \frac{J^{d-4}}{\tilde T^d \zeta_d} = (2\sqrt{x})^{d-4}
(1+x)^{d-2m} \Big(1+x/x^{(m)}_\star \Big)^{-d}  \label{jtrat}
\\ &\zeta_d \equiv \Big(\frac{V_p V(S^{d-1}) }{16 \pi G}
\Big)^{d-4} h^{(d-2)^2} &
\end{eqnarray}
\end{subequations}
where $x^{(m)}_\star$ is defined in \eqref{critvalgce}. The
numerical values of  the relevant ratio  $\tilde j_c $
 on the boundary are also listed in the
Table \ref{tabcrit}. Note that, in analogy with the quantity $\tilde \omega$
relevant for the grand canonical ensemble, in this case the
boundary of stability occurs also precisely at the maximum of the
ratio $\tilde j$ in \eqref{jtrat}. One can also easily obtain the 
corresponding critical
values of $\tilde \omega$ using \eqref{mgentom} and the critical values
$\hat x_c^{(m)}$. 
In parallel with the grand
canonical ensemble the phase diagram for the cases with a boundary
of stability consists again of a stable and unstable sheet. It
would be interesting to examine the possibility of phase mixing
along the lines described in the previous subsection.

\subsection{Critical exponents
\label{subseccritexp} }

We conclude this section with a general analysis of the critical
exponents in both the ensembles for the case of one non-zero
angular momentum. To this end, we note that besides the specific
heats \eqref{com} and \eqref{cj}, one also has the response
functions
\begin{subequations}
\begin{equation}
\chi_T = \Big( \frac{\partial J}{\partial \Omega} \Big)_T =
\frac{V_p V(S^{d-1}) h^{d-2} }{8 \pi G} r_H^2 (1+x)^2
\frac{\Delta_{TJ}}{\Delta_{T\Omega}}
\end{equation}
\begin{equation}
\alpha_\Omega = \Big( \frac{\partial J}{\partial T} \Big)_\Omega =
\frac{V_p V(S^{d-1}) h^{d-2} }{G}  r_H^2 \sqrt{x} \
 \frac{2-(d-4)x }{\Delta_{T\Omega}}
\end{equation}
\begin{equation}
\alpha_J = \Big( \frac{\partial \Omega}{\partial T} \Big)_J =
- \frac{8 \pi \sqrt{x} } { (1+x)^2} \frac{2-(d-4)x }{\Delta_{T J}}
\end{equation}
\end{subequations}
where $\chi_T$ is the isothermal capacitance.  The following
discussion pertains to the cases in which a boundary of stability
was found in the one angular momentum case, i.e. $d \geq 5$ in the
grand canonical ensemble, and $d=3,5,6,7$ in the canonical
ensemble.

Starting with the grand canonical ensemble,
 we consider a point $(T_c,\Omega_c)$ on the
boundary of stability\footnote{The boundary of stability does not
have any special point other than $(T=0,\Omega=0)$ so we take a
generic point different from that.}.
Following a similar analysis
as in \cite{Cai:1998ji}, we show that this point behaves as a
critical point in ordinary thermodynamics.
The stable region has $T \geq T_c$ and $\Omega \leq \Omega_c$, so
we define the quantities
\begin{equation}
\epsilon_T = \frac{T-T_c}{T_c},\ \ \epsilon_\Omega =
\frac{\Omega_c-\Omega}{\Omega_c}
\end{equation}
and consider a function $f(T,\Omega)$ near the point \(
(T_c,\Omega_c) \). The critical exponents $n_T$ and $n_\Omega$ for
$f(T,\Omega)$ are then defined as
\begin{subequations}
\begin{equation}
\label{ntdef}
 n_T = - \lim_{\epsilon_T \rightarrow 0} \frac{\ln
f}{\ln \epsilon_T} \Big|_{\epsilon_\Omega=0} = - \lim_{\epsilon_T
\rightarrow 0} \frac{d \ln f|_{\epsilon_\Omega=0}}{d \ln
\epsilon_T}
\end{equation}
\begin{equation}
n_\Omega = - \lim_{\epsilon_\Omega \rightarrow 0} \frac{\ln f}{\ln
\epsilon_\Omega} \Big|_{\epsilon_T=0} = - \lim_{\epsilon_\Omega
\rightarrow 0} \frac{d \ln f|_{\epsilon_T=0}}{d \ln
\epsilon_\Omega}
\end{equation}
\end{subequations}
We assume that the function satisfies
\begin{equation}
\label{funcform}
f(T,\Omega_c) = \frac{g(T)}{h_T (x)} \sp
f(T_c,\Omega) = \frac{\tilde g(\Omega )}{h_\Omega (x)}
\end{equation}
where $g(T_c)$
and $\tilde g(\Omega_c)$
are finite and different from zero and both $h_T$ and $h_\Omega$
satisfy  \( h |_{x=x_c}=0 \) and \( \frac{dh}{dx} |_{x=x_c} \neq 0 \). This is indeed
true for the response functions $C_\Omega$,
$\chi_T$, $\alpha_\Omega$ and the quantities $(S-
S_c)^{-1}$ and $(J-J_c)^{-1}$.

We first approach the critical point by varying the temperature,
and hence  put \( \Omega = \Omega_c \). Using (\ref{gentom}) and
(\ref{crittom}) one obtains
\begin{equation}
\epsilon_T (x) = \frac{1}{2}
 \sqrt{ \frac{x_c}{x} } \left(1+\frac{x}{x_c} \right) -1
\end{equation}
and substituting in \eqref{ntdef} one finds
\begin{eqnarray}
n_T &=& - \lim_{\epsilon_T \rightarrow 0} \frac{d \ln
f|_{\epsilon_T=0}}{d \ln \epsilon_T}  = - \lim_{\epsilon_T
\rightarrow 0} \frac{1}{f} \frac{d f}{d \ln \epsilon_T} = -
\lim_{x \rightarrow x_c} \frac{\epsilon_T}{f} \frac{d f}{d x}
\Big( \frac{d \epsilon_T}{dx} \Big)^{-1} \nn \\ &=& - \lim_{x
\rightarrow x_c} \frac{\epsilon_T}{h_T} \frac{d h_T}{d x} \Big(
\frac{d \epsilon_T}{dx} \Big)^{-1} = \frac{1}{2}
\end{eqnarray}
Here, the last step follows from \( h_T  |_{x=x_c} = 0\), \(  \frac{dh_T}{dx} |_{x=x_c} \neq 0\),
\( \frac{d\epsilon_T}{dx} |_{x=x_c} = 0\) and \( \frac{d^2
\epsilon_T}{dx^2} |_{x=x_c} \neq 0\).
On the other hand, approaching the critical line by varying
$\Omega$, we need to put \( T=T_c \) and have
\begin{equation}
\epsilon_\Omega (x) = 1- 2 \sqrt{ \frac{x}{x_c} } \frac{1}{ 1 + x/x_c}
\end{equation}
Since \( \frac{d\epsilon_\Omega}{dx} |_{x=x_c} = 0\) and \(
\frac{d^2 \epsilon_\Omega}{dx^2} |_{x=x_c} \neq 0\) we also find
that \( n_\Omega = \frac{1}{2} \).

Since each of  the functions $C_\Omega$, $\chi_T$,
$\alpha_\Omega$, $(S-S_c)^{-1}$ and $(J-J_c)^{-1}$ is of  the form
(\ref{funcform}), one immediately concludes that for each of these
quantities the critical exponents is equal to $\frac{1}{2}$. The
common value $\frac{1}{2}$, which was earlier found
\cite{Cai:1998ji} for the non-dilatonic branes\footnote{These
critical exponents should be related to the corresponding
exponents in correlation functions of the field theory. In the
field theory analysis for the D3-brane  case \cite{Gubser:1998jb}
no agreement was found, but a mean field treatment was suggested
to cure this discrepancy.}, apparently persists and this value has
been shown to be in agreement with scaling laws in statistical
physics \cite{Cai:1998ji}.

The critical analysis in the canonical ensemble proceeds along the
same lines. In this case we consider a point $(T_c,J_c)$ on the
boundary of stability. Repeating the analysis above essentially
with the replacement $\Omega \rightarrow J$, and using the fact
that  $C_J$, $\alpha_J$, $(S-S_c)^{-1}$ and
$(\Omega-\Omega_c)^{-1}$ are all of the form (\ref{funcform}) (with
$\Omega \rightarrow J$), it
follows that also in this case all  critical exponents are \(
\frac{1}{2} \).


\section{Field theory analysis
\label{secfieldtheory} }

\newcommand{\li}{\mathrm{Li}}

In this section we consider the quantum field theories living on
the D and M-branes in the limit where they are
free field theories. Using the ideal gas approximation we compute
in Section \ref{subsecweakfe} the free energies with non-zero
R-voltage under the R-symmetry. We do this to compare the
thermodynamic behaviour in the weak coupling limit with the strong
coupling limit. In Section \ref{subsecinterpo} we discuss the
interpolation between weak and strong coupling, while in Section
\ref{subsecweaktherm} we find the stability behaviour at weak
coupling and compare this to the strong coupling limit.

\subsection{The free energy for weakly coupled field theory
\label{subsecweakfe} }

In this section we calculate the free energies for the
extremely weakly coupled limit of the dual field theories
extending the regularization method used in
\cite{Gubser:1998jb,Cvetic:1999rb} for the D3, M2 and M5-brane.

We start by writing the free energy with all R-charge voltages \(
\{ \Omega_i \} \) turned off. As we shall see, this depends only
on the spatial dimension $p$ of the field theory, and on the
number of massless bosonic and fermionic degrees of freedom. In
particular, the field theories that we consider have 16
supercharges so that for $N=1$ these theories have 8 bosonic and 8
fermionic degrees of freedom. Using the ideal gas approximation,
where particles are assumed to have negligible interaction, we get
the free energy\footnote{For a confining theory, it is understood
that the temperature is above the confining temperature.}
\begin{equation}
F = T V_p \int \frac{d^p q}{(2\pi)^p}
\left[ 8 \log \Big( 1 - e^{-\beta |q|} \Big)
- 8 \log \Big( 1 + e^{-\beta |q|} \Big) \right]
= - k_p V_p T^{p+1}
\end{equation}
with
\begin{equation}
k_p = 2^{4-p} (2 - 2^{-p})
\frac{(p-1)! }{\Gamma(p/2) \pi^{p/2}} \zeta(p+1)
\end{equation}
where $p \geq 1$.

If we consider non-zero R-voltage, we must replace
\( \beta |q| \) with \( \beta |q| + \beta \sum_{i=1}^n \alpha_i \Omega_i \)
in the partition function, where
\( \vec{\alpha} = {(\alpha_1,\alpha_2,...,\alpha_n)} \)
is the $SO(d)$ weight vector  of the particle.
The resulting free energy is
\begin{equation}
\label{free1}
F = T V_p \int \frac{d^p q}{(2\pi)^p} \sum_{\vec{\alpha}} s_{\vec{\alpha}}
\log \left[ 1 - s_{\vec{\alpha}} \exp \Big( - \beta |q|
- \beta \sum_{i=1}^n \alpha_i \Omega_i \Big) \right]
\end{equation}
where \( \vec{\alpha} \) runs over the 16 different particles and
\( s_{\vec{\alpha}} \) is \( +1 \) for bosons and \( -1 \) for fermions.
The weights \( \vec{\alpha} \) for the different \( SO(d) \) R-charge groups
are listed in Table \ref{tabweights}. Note that the weights for
\( SO(2n) \) and \( SO(2n+1) \) are the same, and that branes
with identical R-symmetry group have the same weights.

\begin{table}
\begin{center}
\begin{tabular}{|c||c|c|}
\hline
$n$  & Bosons        & Fermions  \\ \hline
1  & $6(0),\ (\pm 1)$   & $4(\pm \frac{1}{2})$     \\ \hline
2  & $4(0),\ (\pm 1 ,0),\ (0,\pm 1)$    &  $2(\pm \frac{1}{2},\pm \frac{1}{2})$    \\ \hline
3  & $2(0),\ (\pm 1 ,0,0),\ (0,\pm 1 ,0),\ (0,0, \pm 1)$   &   $(\pm \frac{1}{2},\pm \frac{1}{2},\pm \frac{1}{2})$   \\ \hline
4  & $(\pm 1 ,0,0,0),\ (0,\pm 1 ,0,0),$   &    $(\pm \frac{1}{2},\pm \frac{1}{2},\pm \frac{1}{2},\pm \frac{1}{2})$  \\
   & $(0,0, \pm 1,0),\ (0,0,0, \pm 1)$  & number of pluses = even \\ \hline
\end{tabular}
\end{center}
\caption{Weights for the 8 bosons and 8 fermions in the four
possible cases labeled by $n=[\frac{d}{2}]$, corresponding to $3
\leq d \leq 8$. Numbers in front of the weights denote the
degeneracy of the spectrum with respect to this weight. In the
$n=4$ case the 8 fermions all have same chirality under the
$SO(8)$. \label{tabweights}   }
\end{table}

The integrals for the 8 bosons in \eqref{free1} are clearly
divergent since \( \beta \Omega_i \) is real. In Ref.
\cite{Gubser:1998jb} it was proposed to perform an analytic
continuation by considering \( \beta \Omega_i \) to be complex, so
that using \eqref{liint} the free energies  \eqref{free1} can be
expressed in terms of polylogarithms,
\begin{equation}
\label{free2}
F = - \frac{\Gamma(p)}{2^{p-1} \pi^{p/2} \Gamma(p/2) } V_p T^{p+1}
\sum_{\vec{\alpha}} \li_{p+1} \left[ s_{\vec{\alpha}}
\exp \Big( - \sum_{i=1}^n \alpha_i \omega_i \Big) \right]
\end{equation}
where \( \omega_i = \beta \Omega_i \). The polylogarithms are not
defined for real numbers greater than one, but in Appendix
\ref{apppoly} we discuss the continuation to this region,
along with some general properties of polylogarithms.

Using the exact functions \( B_n(x) \) and \( F_n(x) \) for \( x
\in \mathbb{R} \) of Appendix \ref{apppoly}, we can in principle
write all the free energies for the different $p$-branes exactly.
To save space, we restrict ourselves to write the energies with
odd $p$ exactly, and write the energies with even $p$ to fourth
order in $\omega_i$. In Section \ref{subsecweaktherm}, however, we
use the fact that all the free energies are known to all orders in
\( \omega_i \).

For the M-branes, it is believed that $N=1$ corresponds to a
free field theory, while for $N>1$ the field theories are interacting.
Thus, for a single M2-brane and M5-brane we have
\begin{subequations}
\label{weakmfe}
\begin{eqnarray}
F_{\mathrm{M2}} &=& - V_2 T^3 \frac{1}{\pi} \left[ 7\zeta(3) -
\frac{1}{2} \sum_{i=1}^4 \log( \omega_i ) \omega_i^2 + \Big(
\frac{1}{2}\log(2) + \frac{3}{4} \Big) \sum_{i=1}^4 \omega_i^2
\right. \nn \\ && \left. + \frac{1}{128} \Big( \sum_{i=1}^4
\omega_i^2 \Big)^2 - \frac{10}{1152} \sum_{i=1}^4 \omega_i^4 +
\frac{1}{16} \omega_1 \omega_2 \omega_3 \omega_4 +
\Ord(\omega_i^6) \right]
\end{eqnarray}
\begin{eqnarray}
F_{\mathrm{M5}} &=& - V_5 T^6 \left[ \frac{\pi^3}{30}
+ \frac{\pi}{24}(\omega_1^2+\omega_2^2)
+ \frac{1}{96\pi}(\omega_1^2+\omega_2^2)^2
+ \frac{1}{48\pi}(\omega_1^4+\omega_2^4) \right.
\nn \\ &&
\left.
+ \frac{1}{1152\pi^3} (\omega_1^2+\omega_2^2)^3
-\frac{1}{288\pi^3} (\omega_1^6+\omega_2^6) \right]
\end{eqnarray}
\end{subequations}
The ideal gas approximation is valid for the D-branes when \(
\lambda = 0 \). In this limit, the free energies for $N$
D$p$-branes take the form
\begin{subequations}
\label{weakdfe}
\begin{equation}
F_{\mathrm{D1}} = - 2 \pi N^2 V_1 T^2
\end{equation}
\begin{eqnarray}
F_{\mathrm{D2}} &=& - N^2 V_2 T^3 \frac{1}{\pi} \left[ 7\zeta(3) -
\frac{1}{2} \sum_{i=1}^3 \log( \omega_i ) \omega_i^2 + \Big(
\frac{1}{2} \log(2) + \frac{3}{4} \Big) \sum_{i=1}^3 \omega_i^2
\right. \nn \\ && \left. + \frac{1}{128} \Big( \sum_{i=1}^3
\omega_i^2 \Big)^2 - \frac{10}{1152} \sum_{i=1}^3 \omega_i^4 +
\Ord(\omega_i^6) \right]
\end{eqnarray}
\begin{equation}
\label{weakD3free}
F_{\mathrm{D3}} = - N^2 V_3 T^4 \left[ \frac{\pi^2}{6}
+ \frac{1}{4} \sum_{i=1}^3 \omega_i^2
+ \frac{1}{32\pi^2} \Big( \sum_{i=1}^3 \omega_i^2 \Big)^2
- \frac{1}{16\pi^2} \sum_{i=1}^3 \omega_i^4 \right]
\end{equation}
\begin{eqnarray}
F_{\mathrm{D4}} &=& - N^2 V_4 T^5 \frac{1}{\pi^2} \left[
\frac{93}{8}\zeta(5) + \frac{21}{16}\zeta(3)
(\omega_1^2+\omega_2^2) + \Big( \frac{25}{192} +
\frac{1}{64}\log(2) \Big) (\omega_1^4 + \omega_2^4) \right. \nn \\
&& \left. + \frac{3}{32}\log(2) \omega_1^2 \omega_2^2 +
\frac{1}{16} \Big( -\log(\omega_1) \omega_1^4 - \log(\omega_2)
\omega_2^4 \Big) + \Ord(\omega_i^6) \right]
\end{eqnarray}
\begin{eqnarray}
F_{\mathrm{D5}} &=& - N^2 V_5 T^6 \left[ \frac{\pi^3}{30}
+ \frac{\pi}{24}(\omega_1^2+\omega_2^2)
+ \frac{1}{96\pi}(\omega_1^2+\omega_2^2)^2
+ \frac{1}{48\pi}(\omega_1^4+\omega_2^4) \right.
\nn \\ &&
\left.
+ \frac{1}{1152\pi^3} (\omega_1^2+\omega_2^2)^3
-\frac{1}{288\pi^3} (\omega_1^6+\omega_2^6) \right]
\end{eqnarray}
\begin{equation}
F_{\mathrm{D6}} = - N^2 V_6 T^7 \frac{1}{\pi^3} \left[
\frac{1905}{64}\zeta(7) + \frac{465}{128}\zeta(5) \omega^2 +
\frac{95}{512}\zeta(3) \omega^4 + \Ord(\omega^5) \right]
\end{equation}
\end{subequations}
%

\subsection{Interpolation between weakly and strongly coupled theories
\label{subsecinterpo} }

While the expressions \eqref{weakmfe} represent the
free energies of the M-branes for $N=1$, the supergravity results
\eqref{strongm2fe} and \eqref{strongm5fe} are the corresponding
free energies in the $N \rightarrow \infty$ limit.
As discussed in \cite{Gubser:1998nz} (without R-voltage $\Omega_i$)
it is expected that there is a smooth interpolating function
$f(N,\{\omega_i\})$
so that the free energy for all $N$ is given by
\begin{equation}
\label{interm}
F_N (T,\{\Omega_i\}) = f(N,\{\omega_i\}) F_{N=1} (T,\{\Omega_i\})
\end{equation}
where \( F_{N=1} (T,\{\Omega_i\}) \) is the free energy for $N=1$,
given in \eqref{weakmfe}. In other words, one can conjecture that
if all higher derivative terms in the effective 11-dimensional
supergravity action were known, and if one could find the spinning
black M-brane solution in this effective action, one could compute
the free energies for all $N$, and in particular the free energies
\eqref{weakmfe} for $N=1$. The status of this conjecture, however,
is not clear since there could very well be a phase transition
obstructing the smooth interpolation to the free theory limit.

Turning to the D-branes, the expressions in \eqref{weakdfe}
represent the free energies for $N \gg 1$ and \( \lambda = 0 \)
(which in particular means that $ \lambda \ll r_H^{3-p} $), while
the supergravity results \eqref{strongdpfe} are valid for $N \gg
1$ and large 't Hooft coupling, \( \lambda \gg r_H^{3-p} \) (for the
D$p$-branes with $p\neq 3$ there is also has an upper bound on
$\lambda$, see Eq. \eqref{geff}). Thus, we can consider the free
energies of D-branes with $N$ fixed but with $\lambda$ varying
between the two limits just described. One can then conjecture
that for fixed $N \gg 1$ there exists a smooth interpolation
function $f(\lambda,T,\{ \Omega_i \} )$ so that we can write
\begin{equation}
\label{interd}
F_\lambda (T,\{\Omega_i\}) = f(\lambda,T,\{ \Omega_i \} )
F_{\lambda=0} (T,\{\Omega_i\})
\end{equation}
where \( F_{\lambda=0} (T,\{\Omega_i\}) \) is the free energy for
\( \lambda = 0 \). Moreover, for the D3-brane  the field theory is
conformal, so that the function  is expected to depend on
dimensionless quantities only, i.e. $f(\lambda,\{ \omega_i \})$.
The possibility of such a smooth interpolation has previously been
discussed in \cite{Gubser:1998nz} for the D3-brane without the
R-voltage. An important first check of this conjecture is the fact
that the free energies of the D-branes in the two limits show the
same $N^2$ factor in front. This implies that only string loop
corrections, which carry factors of $\frac{1}{N^2}$ would modify
this behavior, and thus do not have to be considered in the large
$N$ limit. Comparing the $\lambda$-dependence on the other hand,
one sees that only for the D3-brane the same form is found in the
two limits. Again, if one knew the higher derivative terms of the
effective action of type II string theory and one could solve the
equations of motion for a spinning black D-brane, one could
presumably find the smooth interpolation between the two limits of
$\lambda$. Again, this conjecture assumes that there is not a
phase transition between weak and strong coupling. This assumption
has been challenged in \cite{Li:1998kd} for the D3-brane.

We will return to this issue in Section \ref{secd3corr}, where we
calculate the leading order correction in $\lambda^{-3/2}$
to the D3-brane free energy, due to the $l_s^6 R^4$ term in
the type IIB effective action.
In Section \ref{subsecweaktherm} we take another path and compare
the thermodynamic behavior of the field theory in the two limits,
including a study of the thermodynamic stability.

\subsection{Stability behavior at weak coupling
\label{subsecweaktherm} }

We analyze the thermodynamic stability of the weakly coupled QFTs
using the free energies \eqref{weakmfe} and \eqref{weakdfe}. For
simplicity, we restrict ourselves to the case of one non-zero
voltage $\Omega_1=\Omega$. As a consequence, branes with equal
number of spatial dimensions $p$ have the same thermodynamics,
since our analysis is not affected by the overall dependence on
$N$.

{} From the Gibbs free energy $F=F(T,\Omega)$ we compute the heat
capacity
\begin{equation}
C_\Omega = - T \Big( \frac{\partial^2 F}{\partial T^2} \Big)_\Omega
\end{equation}
and in the cases we consider, one can check that \( C_\Omega \) is
always positive. Instead we can extract the stability behaviour by
considering the Hessian matrix
\begin{equation}
\label{eqhes}
\hes (F) = \left( \begin{array}{cc} \frac{\partial^2 F}{\partial T^2}
& \frac{\partial^2 F}{\partial T \partial \Omega} \\
\frac{\partial^2 F}{\partial T \partial \Omega}
& \frac{\partial^2 F}{\partial \Omega^2} \end{array} \right)
\end{equation}
of the free energy $F=F(T,\Omega)$. For a stable point, the
Hessian \eqref{eqhes} should be negative definite.
Since \( C_\Omega \) is positive for \( \Omega = 0 \), the
boundary of stability is reached  when one of the eigenvalues of
the Hessian \eqref{eqhes} changes sign. Since there are no
singularities this occurs when $\det(\hes (F) ) = 0$, so that the
boundary of stability is characterized by a certain critical value
of $\omega = \Omega/T$. The results of the analysis for the
various values of $p$ are given in Table \ref{tabweakcrit}. To
analyze the cases of even $p$, we use that we know the free energy
to all orders in $\omega$. Thus, one can take an appropriate
number of terms in order to ensure that the value of $\omega$ that
one finds has the required accuracy.

\begin{table}
\begin{center}
\begin{tabular}{|c||c|}
\hline
$p$  & $\omega_c$   \\ \hline
1    & Stable     \\ \hline
2    & $1.5404$   \\ \hline
3    & $2.4713$   \\ \hline
4    & $3.3131$   \\ \hline
5    & $4.1458$   \\ \hline
6    & $4.9948$   \\ \hline
\end{tabular}
\end{center}
\caption{The boundary of stability for the various $p$-branes in
the weakly coupled field theory limit.  \label{tabweakcrit}   }
\end{table}

Considering Table \ref{tabweakcrit} we see that most branes have a
boundary of stability at a certain value of $\omega$, as also seen
in the stability analysis of Section \ref{subsecgrandcan}. We also
remark that all the values of $\omega$ in Table \ref{tabweakcrit}
have the same orders of magnitude as the values in Table
\ref{tabcrit}, and thus it seems plausible that the conjectured
interpolation between the two limits of the QFTs described in
Section \ref{subsecweakfe} should connect the values of $\omega$.
Table \ref{tabcompomega} summarizes  the values of $\omega$ for
the weak and strong coupling limits of the D, and M-branes, and
for the D2, D3, D4, M2 and M5-branes the critical values of
$\omega$ in the two limits are seen to be remarkably close. We
also note that $\omega$ is increasing with $p$ in both limits.

\begin{table}
\begin{center}
\begin{tabular}{|c||c|c|}
\hline
Brane  & $\omega_{\mathrm{weak}}$ & $\omega_{\mathrm{strong}}$ \\ \hline
D1 & Stable   & $1.2825$ \\ \hline
D2     & $1.5404$ & $1.6223$ \\ \hline
D3     & $2.4713$ & $2.2214$ \\ \hline
D4     & $3.3131$ & $3.6276$ \\ \hline
D5 & $4.1458$ & Not defined  \\ \hline
D6     & $4.9948$ & Unstable \\ \hline
M2     & $1.5404$ & $1.2825$ \\ \hline
M5     & $4.1458$ & $3.6276$ \\ \hline
\end{tabular}
\end{center}
\caption{Comparison between the boundaries of stability
for the type II D$p$-branes n the weak
and strong coupling limits of $\lambda$
and for the M-branes
in the $N=1$ and $N\rightarrow \infty$ limits.
\label{tabcompomega}   }
\end{table}

The D1-brane and D6-brane, however, are seen from Table
\ref{tabcompomega} to have qualitatively different stability
behaviour in the weak and strong coupling limits, so that in this
case the interpolation should somehow create or destroy a boundary
of stability at some special point between the two limits. If we
for definiteness think about the QFT living on $N$ D1-branes on
top of each other, we see that for $\lambda = 0$ it is stable,
also with  R-voltage turned on, while for $\lambda$ large it should
exhibit a boundary of stability. Thus, at some value of $\lambda$ the
QFT makes a transition from being completely stable to being
potentially unstable. It would be interesting to study how this
mechanism works in detail.

For the D5-branes we also have completely different
qualitative behaviour. At weak coupling we can vary the
thermodynamic parameters \( T \) and \( \{ \Omega_i \} \) freely,
while at strong coupling, they are constrained (see Section
\ref{subsecgrandcan}). Thus, somehow the phase space must expand
as one moves away from strong coupling. One could also try to
study this phenomenon for non-rotating branes, here the
temperature is constant at strong coupling.

If we instead consider the canonical ensemble, with variables $T$
and $J$, one must consider the heat capacity
\begin{equation}
C_J = T \Big( \frac{\partial S}{\partial T} \Big)_J
= T \det( \hes (F)  )
\left[ \Big( \frac{\partial J}{\partial \Omega} \Big)_T \right]^{-1}
\end{equation}
Thus, we see that $C_J$ is zero whenever $\det(\hes (F) )$ is,
i.e. the canonical ensemble has the same stability behaviour as
the grand canonical ensemble.
This  result  should not be surprising since we in fact have used
standard statistical physics to derive our thermodynamic
relations, so that it is expected that general results,  
such as the equivalence of
ensembles, should hold. Nevertheless, it would be interesting to
test this by computing corrections to the stability behaviour from
the weakly coupled field theory side, or the supergravity side,
since the results of Section \ref{seccritbeh} show that the
thermodynamic ensembles are not equivalent in the strongly coupled
large $N$ limit. If we consider a D-brane, the expectation would
be that the boundaries of stability in the two different ensembles
start for $\lambda=0$ at the same value, move away from each other
as $\lambda$ increases and finally reach  the values given in
Section \ref{seccritbeh}.

Finally we note that, with respect to the critical exponents there is
also a qualitative difference between the
weak and strong coupling limit of the QFT.
As discussed in Section \ref{subseccritexp} the heat capacities
$C_\Omega$ and $C_J$ both behave as $1/\sqrt{T-T_c}$ near the boundary
of stability. But, in the weak coupling limit one can easily check
that $C_\Omega$ and $C_J$ both behave as $(T-T_c)^\alpha$ with
$\alpha \geq 0$, $\alpha$ being different for the two heat capacities.
Another way to see this, is to note that while the heat capacities
in the strong coupling limit have singularities on the boundary of stability,
they are continuous in the weak coupling limit.

In conclusion we see that there are many similarities between the
thermodynamics at weak and at strong coupling (or small and large
$N$ for the M-branes), but also important qualitative differences
that are non-trivial to connect by an interpolation between the
two limits. In the next section we make the first step towards a
quantitative understanding of this conjectured interpolation for
spinning branes.

\section{Free energies from the supergravity action}
\label{secfefromaction}

It is well known that one can obtain the free energy of a field
theory by Wick rotation and evaluating the path integral partition
function with the boundary condition that time is periodic with
the inverse temperature as period. For general relativity, this
method has been applied in order to compute the free energy
of a black hole, but, with some difficulty since one need to think
of ways to circumvent the problems of quantizing the gravitational
field. In Anti-de Sitter space though, this has proved
surprisingly easy since one can just evaluate the action on the
background geometry \cite{Hawking:1983dh}. Moreover, the free
energy for the near-horizon limit of the D3, M2 and M5-brane has
been reproduced with this method \cite{Witten:1998zw}, which is
not surprising since these branes all have Anti-de Sitter space
times a sphere as their near-horizon geometry.

In Section \ref{secfreeenergy} we will extend these results to
spinning branes in the near-horizon limit, showing that 
we are able to reproduce the free
energy \eqref{nhgibbs} found in Section \ref{nhtherm}. In Section
\ref{secd3corr} we  compute
the first correction in
$1/\lambda$ to the free energy of the spinning D3-brane
found in Section \ref{secfreeenergy}.  This
will then be used to test the conjecture that there exists a
smooth interpolating function between $\lambda=0$ and $\lambda
=\infty$, as discussed in Section \ref{subsecinterpo}.

\subsection{Free energies from the low-energy effective action
\label{secfreeenergy}}

The Euclidean low-energy supergravity action is 
\begin{equation}
\label{euclact}
I_E = I_E^{\mathrm{bulk}} + I_E^{\mathrm{bd}}
\end{equation}
where the bulk term is   given by  
\begin{equation}
\label{bulkaction}
I_E^{\mathrm{bulk}} = - \frac{1}{16\pi G} \int_{\cal{M}} d^D x \sqrt{g}
\Big( R-\frac{1}{2} \partial_\mu
\phi \partial^\mu \phi - \frac{1}{2(p+2)!} e^{a\phi} F_{p+2}^2 \Big)
\end{equation}
and the boundary term is \cite{Gibbons:1977ue}  
\begin{equation}
\label{boundaryterm}
I_E^{\mathrm{bd}} = - \frac{1}{8\pi G}
\int_{\partial {\cal{M}} } d^{D-1} x \sqrt{h} K
\end{equation}
with $h_{\mu \nu}$ the boundary metric and $K$ the
extrinsic curvature. In this section we obtain the 
free energy
\eqref{nhgibbs} as $I_E/ \beta$, where $\beta=1/T$ is the inverse
temperature and $I_E$ is the regularized value of the
Euclidean action \eqref{euclact}    
evaluated on the Wick rotated spinning $p$-brane solution in the
near-horizon limit.  
We restrict
ourselves to one non-zero angular momentum, since we expect that
because the free energy \eqref{nhgibbs} is independent of the
angular momentum, more non-zero angular momenta will not alter our
final result.

In Appendix \ref{appeuclnhsol} we perform a Wick rotation of the
near-horizon solution \eqref{nhsol} in the presence of one
non-zero angular momentum, to obtain the Euclidean spinning brane
solution \eqref{nheucl}. Starting with the bulk term \eqref{bulkaction},  
we  substitute \eqref{nheucl} and integrate over the time $\tau$ and the
angles to arrive at the following general expression
\begin{equation}
\label{angint}
\frac{L (r) }{\beta} =
\frac{V_p V(S^{d-1})}{16 \pi G} \frac{(d-2)^3}{D-2}
r^{d-3}\left[ 1 + \sum_{s=1}^{\infty}
\left(\tilde v_s + \tilde w_s \left( \frac{r_0}{r} \right)^{d-2} \right)
 \Big( \frac{\eul}{r} \Big)^{2s}  \right]
\end{equation}
where the $\beta=1/T$ factor is the period of the time $\tau$.
Here the coefficients $\tilde v$ and $\tilde w$
can be computed in principle through
any desired order for any brane solution.

To evaluate the final integral over $r$ we need to introduce a regularization
method along the lines of \cite{Witten:1998zw,Gubser:1998nz}.
In this prescription we first perform
the integral up to a cutoff radius $\rma$ and subtract the contribution
of the extremal brane with a temperature equal to the original brane at
the cutoff radius $\rma$.
Thus, integrating the expression \eqref{angint} from the horizon radius $r_H$
to the cutoff radius $\rma$ we arrive at
\begin{equation}
\label{rint}
\left.
\frac{I_E^{\rm bulk} }{\beta} =
\frac{V_p V(S^{d-1}) }{16 \pi G} \frac{(d-2)^2}{D-2}
r^{d-2}\left[ 1 + \sum_{s=1}^{\infty}
\left( v_s +  w_s \left( \frac{r_0}{r} \right)^{d-2} \right)
 \Big( \frac{\eul}{r} \Big)^{2s } \right] \right\vert_{r_H}^{\rma}
\end{equation}
In particular, for the near-horizon spinning solutions in 10-dimensional
 type
II string theory and 11-dimensional  M-theory, one finds by explicit
evaluation that the expansion coefficients $v_s, w_s$ can be uniformly
written \footnote{We have checked these relations up to fourth
order in $(\tilde l/r)^2$ for all 10-dimensional and 11-dimensional
 brane solutions, and we believe that they are generally valid. Using these
values one can also write a closed form expression for
\eqref{rint} in terms of logarithms.} as
\begin{subequations}
\begin{equation}
 v_1 = -\frac{d-2}{d} \sp  v_s =
-\frac{4}{(2s+ d-4 )(2s + d-2) } \sp s \geq 2
\end{equation}
\begin{equation}
 w_s  = -\frac{2}{2s+ d-2 } \sp s \geq 1
\end{equation}
\end{subequations}
with $d$ the transverse dimension.
Note that these coefficients satisfy the recursive relations
\begin{equation}
\label{recrel}
v_s =  w_{s-1} - w_{s} \sp  v_0 = 1 \sp  w_0 = -1
\end{equation}
the importance of which will become apparent below.
Continuing with \eqref{rint} we find after some algebra that
\begin{eqnarray}
\label{exp1}
\frac{I_E^{\rm bulk} }{\beta} &=&
\frac{V_p V(S^{d-1})}{16 \pi G} \frac{(d-2)^2}{D-2}
\left[ \rma^{d-2} - r_0^{d-2}  
+ \sum_{s=1}^{\infty} \left(v_s  \rma^{d-2}
+ w_s r_0^{d-2} \right) \left( \frac{\eul}{\rma} \right)^{2s} 
\right.
\nn \\ &&
\left. 
- \sum_{s=0}^{\infty} ( v_s r_H^{d-2} + w_s r_0^{d-2} )
 \left( \frac{\eul}{r_H} \right)^{2s } \right]
\end{eqnarray}
Using the relation \eqref{horizon}
to write $r_0^{d-2} = r_H^{d-2} ( 1- (\eul/r_H)^2)$
the recursion relation \eqref{recrel} implies that the 
last term in \eqref{exp1} cancels, giving  
\begin{equation}
\label{exp2}
\frac{I_E^{\rm bulk} }{\beta} 
=\frac{V_p V(S^{d-1})}{16 \pi G} \frac{(d-2)^2}{D-2}
\left[ \rma^{d-2} - r_0^{d-2}  + \sum_{s=1}^{\infty}
(v_s  \rma^{d-2}
+ w_s r_0^{d-2} ) \left( \frac{\eul}{\rma} \right)^{2s} \right]
\end{equation}
The regularized bulk contribution to the free energy is
\begin{equation}
\label{regfree0}
F_{\rm bulk} = \lim_{\rma \rightarrow \infty} \left[
\frac{I_E^{\rm bulk} }{\beta} - \left. \frac{I_E^{\rm bulk}}{\beta'} \right\vert_{r_0=0} \right]
= \lim_{\rma \rightarrow \infty} \left[
\frac{I_E^{\rm bulk} }{\beta} - \left. \frac{\beta}{\beta'} 
\frac{I_E^{\rm bulk}}{\beta} \right\vert_{r_0=0} \right]
\end{equation}
where the ratio of the temperatures is given by
\begin{equation}
\label{tempratio}
\frac{\beta}{\beta'} = f^{1/2}\vert_{r=\rma} = 1 - \frac{1}{2}
\left( \frac{r_0}{\rma} \right)^{d-2} + {\Ord} (\rma^{-d} )
\end{equation}
We note that this expression is meaningful since there is no
dependence on the angles to order ${\Ord} (\rma^{-d})$.
Substituting \eqref{exp2} and \eqref{tempratio} in
\eqref{regfree0} we then find after taking the limit the result
\begin{equation}
\label{regfree}
F_{\rm bulk} =-
\frac{V_p V(S^{d-1})}{16 \pi G} \frac{(d-2)^2}{2(D-2)} r_0^{d-2}
\end{equation}
To find the boundary contribution we note that there are two boundaries,
 at \( r=r_H \) and \( r=\rma \) respectively. The boundary
action \eqref{boundaryterm} then gives\footnote{We thank J. Correia for 
useful discussions about this computation.}  
\begin{eqnarray}
\label{bdact}
\frac{I_E^{\mathrm{bd}}}{\beta}
&=&  - \frac{1}{\beta}  \frac{1}{8 \pi G} \int_{ \partial {\cal{M}}}
d^{D-1} x \Big(\partial_r (\sqrt{g}\sqrt{g^{rr}})\Big)
\sqrt{g^{rr}}
\nn \\
&=& \frac{V_p V(S^{d-1}) }{16 \pi G}
\left[ \left( \frac{(p+1)(d-2)}{D-2} - 2(d-1) \right)
( \rma^{d-2} - r_0^{d-2} ) \right. 
\nn \\  && \left. \phantom{ \frac{(p+1)(d-2)}{D-2} }  
- (d-2) r_0^{d-2}  \right]  \left(1 +  \Ord (\rma^{-2}) \right)   
\end{eqnarray}                               
where we remark that only the boundary at \( r=\rma  \)
contributes. We note that here the $r_0$-dependent terms are either written
explicitly or are of order $\Ord (\rma^{-2})$.    
From \eqref{bdact} and \eqref{tempratio} one then
obtains the regularized boundary contribution to the free energy
\begin{equation}
\label{Fbd}
F_{\mathrm{bd}} = \lim_{\rma \rightarrow \infty} \left[
\frac{I_E^{\mathrm{bd}}}{\beta}
- \frac{\beta}{\beta'} \frac{I_E^{\mathrm{bd} }}{\beta}
\Big|_{r_0=0} \right]
= - \frac{V_p V(S^{d-1}) }{16 \pi G}
\left[  \frac{(p+1)(d-2)}{2(D-2)} - 1 \right] r_0^{d-2}
\end{equation}
which we note vanishes for non-dilatonic branes, as seen using
\eqref{aeq}. Adding the two free energy contributions
\eqref{regfree} and \eqref{Fbd} we get
\begin{equation}
F = F_{\mathrm{bulk}} + F_{\mathrm{bd}}
= - \frac{V_p V(S^{d-1}) }{16 \pi G}
 \frac{d-4}{2}  r_0^{d-2}
\end{equation}                  
which precisely reproduces the thermodynamically computed Gibbs
free energy \eqref{nhgibbs}.     
This fact will be used implicitly when we calculate string
corrections to the free energy of the spinning D3-brane in Section
\ref{secd3corr}.

\subsection{Corrections from higher derivative terms
\label{secd3corr} }

In this section we  test the conjecture that there exists smooth
interpolation functions between strong and weak 't Hooft coupling
$\lambda=g_{\mathrm{YM}}^2 N$ for the D-branes, as discussed in
Section \ref{subsecweakfe}. The idea is to compute the correction
to the free energy from the \( l_s^6 R^4 \) term in type II string
theory since this gives us the first correction in \( 1/\lambda \).

We restrict ourselves to the case of the D3-brane, since the
constant dilaton for the non-corrected solution makes the
computation considerably easier.
The extremal non-rotating D3-brane has the
geometry $AdS_5 \times S^5$ in the near-horizon limit and the dual
field theory is the N=4 $D=4$ SYM \cite{Maldacena:1997re}. For the
spinning D3-brane the dual field theory is N=4 $D=4$ SYM at finite
temperature with the R-voltage turned on. The free energy of the
spinning D3-brane in the strong and weakly coupled limits has
previously been discussed in
\cite{Gubser:1998jb,Kraus:1998hv,Cvetic:1999rb}.

We furthermore restrict ourselves to one non-zero angular momentum only
but the methods we use can easily be extended to more angular momenta.
To simplify the computations, we work in the limit
\( \omega \ll \pi \)
with \( \omega = \Omega / T \). This corresponds to the
limit \( l \ll r_0 \).
We develop all series in $\omega$ to order $\omega^4$, with the next
corrections coming from a $\omega^6$ term. With this, our results are accurate
for $\omega < 1 $ up to about $1\%$.

Thus, we will  test the interpolation between the free energy
\begin{equation}
\label{weakD3fe}
F_{\lambda=0} (T,\Omega) = - N^2 V_3 T^4 \left( \frac{\pi^2}{6}
+ \frac{1}{4} \omega^2 - \frac{1}{32\pi^2} \omega^4 \right)
\end{equation}
for weak coupling, obtained from \eqref{weakD3free}, and the free
energy
\begin{equation}
\label{strongD3fe}
F_{\lambda=\infty} (T,\Omega) = - N^2 V_3 T^4 \left( \frac{\pi^2}{8}
+ \frac{1}{8} \omega^2 + \frac{1}{16\pi^2} \omega^4 + \Ord ( \omega^6 )
\right)
\end{equation}
for strong coupling, obtained from \eqref{strongdpfe}.
In this case we can write the interpolation conjecture \eqref{interd} as
\begin{equation}
\label{interD3}
F_\lambda(T,\Omega)  = f(\lambda,\omega) F_{\lambda=0}(T,\Omega)
\end{equation}
where $f=f(\lambda,\omega)$ is the interpolation function.
To zeroth order in \( 1/\lambda \) we thus have
\begin{equation}
\label{flambdainf}
f(\lambda=\infty,\omega)
= \frac{3}{4} - \frac{3}{8\pi^2} \omega^2 + \frac{69}{64\pi^4}\omega^4
+ \Ord ( \omega^6 )
\end{equation}
Comparing \eqref{weakD3fe} and \eqref{strongD3fe} we see that we should expect
$f(\lambda,\omega)$ to be smaller than one, and we also
expect it to be decreasing with $\lambda$, for fixed $\omega < 1$,
since
\begin{equation}
\label{inequal}
-F_{\lambda=\infty} < -F_{\lambda=0} \mbox{ for } \omega < 1
\end{equation}
i.e. the absolute value of the free energy for $\lambda=\infty$ is
less than the one for $\lambda=0$ for $\omega < 1$. In Ref.
\cite{Gubser:1998nz} the interpolation function
$f(\lambda,\omega)$ was studied for $\omega=0$, and it was found
that
\begin{equation}
\label{zerofcomp}
f(\lambda,0) = \frac{3}{4} + \frac{45}{32}
\zeta(3)  (2 \lambda)^{-3/2}  + \ldots
\end{equation}
by computation of the correction from the \( l_s^6 R^4 \) term in
type IIB string theory. The computation \eqref{zerofcomp} clearly
supports the conjecture that there is a monotonous smooth
interpolation function, since the \( \lambda^{-3/2} \) correction
is positive\footnote{On the weak coupling side, a two loop
calculation \cite{Fotopoulos:1998es} has shown that the leading
correction in $\lambda$ is negative, giving further evidence for
the interpolation conjecture. In \cite{Kim:1999sg} further
corrections in $\lambda$ from the weak coupling side are
considered and also found to support the interpolation
conjecture.}.

As previously stated, the higher derivative correction term\footnote{For 
dual field theories with a smaller amount of
supersymmetry, Refs.\cite{Nojiri:1999uh,Nojiri:1999ji} consider
analogous higher derivative corrections to obtain the modification
of the thermodynamics.}
in the supergravity action for type IIB string theory
that we want to consider is the \( l_s^6 R^4 \) term.
In the Euclidean case, the term is given in the Einstein frame by
\begin{equation}
\label{coract}
\delta I_E = - \frac{1}{16 \pi G} \int d^{10} x \sqrt{g} \, \gamma
 e^{-\frac{3}{2} \phi} W
\end{equation}
with \( \gamma = \frac{1}{8} \zeta (3) l_s^6 \) and
\begin{eqnarray}
W &=& C^{\mu_1 \mu_2  \mu_3 \mu_4} C_{\nu_1  \mu_2 \mu_3 \nu_4}
C_{\mu_1}{}^{\nu_2 \nu_3 \nu_1} C^{\nu_4}{}_{\nu_2 \nu_3 \mu_4}
\nn \\ &&
+ \frac{1}{2}
 C^{\mu_1 \mu_2  \mu_3 \mu_4} C_{\nu_1  \nu_4  \mu_3 \mu_4}
C_{\mu_1}{}^{\nu_2 \nu_3 \nu_1} C^{\nu_4}{}_{\nu_2 \nu_3 \mu_2}
\end{eqnarray}
where \( C_{\mu \nu \rho \sigma} \) is the Weyl tensor. In the
near-horizon limit \eqref{rescal} with \( \ell = l_s \rightarrow
0 \) we have the same term \eqref{coract} in terms of the rescaled
quantities, but with \( \gamma \) rescaled to
\begin{equation}
\gamma = \frac{1}{8} \zeta(3)
\end{equation}
From \eqref{dprel} we furthermore have the relations
\begin{equation}
\label{D3scal}
 h^4 = 2 \lambda \sp \frac{V(S^5) h^8}{16 \pi G} =
\frac{N^2}{8\pi^2} \end{equation}

String theory admits two different kinds of expansions, the loop
expansion in \( g_s \) and the derivative expansion in
 \(\alpha' = l_s^2 \). In particular, for the type IIB $R^4$ term
 there is also  a one-loop term of the
  form \( g_s^2 l_s^6
R^4 \), and an infinite sum of D-instanton corrections.
 Through the AdS/CFT correspondence this translates into
 a \( 1/N \) and  \(
1/\lambda \) expansion (see e.g. \cite{Banks:1998nr} which also discusses
the instanton corrections). The \( l_s^6 R^4 \) tree-level term
becomes then a \( \lambda^{-3/2} R^4 \) term, while the \( g_s^2
l_s^6 R^4 \) one loop term becomes an \( N^{-2} \lambda^{1/2} R^4
\) term. The \( N^{-2} \lambda^{1/2} R^4 \) term is clearly not
interesting for this computation, since we want to keep $N$ fixed
and large. It is also subleading since we keep \( g_s \) small in
the limit we consider.

We now compute the \( 1/\lambda \) correction to the free energy
\eqref{strongD3fe} by inserting the non-corrected Wick rotated
solution \eqref{euclmet} for the Euclidean spinning D3-brane in
the near-horizon limit, into the higher derivative term
\eqref{coract} as the background geometry. Substituting the
solution we find for the first three terms\footnote{We have
obtained the exact result but refrain from giving this rather
lengthy expression.} in a weak angular momentum expansion
\begin{multline}
W = \frac{180}{h^8}  \left( \frac{r_0}{r} \right)^{16} \left[ 1 +
 \frac{2}{3}    \left[ 10 \cos^2 \theta
+ (4 - 5\cos^2 \theta )
 \left( \frac{r}{r_0} \right)^{4} \right] \left( \frac{\eul}{r} \right)^2
\right.
\\
+ \frac{1}{120}   \left[ 3617 \cos^4 \theta
+ ( -3032 \cos^2 \theta  + 2288 ) \cos^2 \theta
\left( \frac{r}{r_0} \right)^{4} \right.
\\
\left. \left.
+ ( 512 - 904  \cos^2 \theta  + 644 \cos^4 \theta )
\left( \frac{r}{r_0} \right)^{8}
\right]
\left( \frac{\eul}{r} \right)^4 + \ldots \right]
\end{multline}
while  the volume element is
\begin{equation}
\sqrt{g} = h^2 r^3 \cos^3 \theta \sin  \theta
\cos \psi_1 \sin  \psi_1 \left[  1- \frac{1}{2} \cos^2 \theta
\left( \frac{\eul}{r} \right)^2
-\frac{1}{8} \cos^4 \theta \left( \frac{\eul}{r} \right)^4 + \ldots \right]
\end{equation}
Substituting this in \eqref{coract}, integrating over the angles, the
world-volume, the Euclidean time $\tau$
from 0 to $\beta$,  and $r$ from $r_H$ to
infinity, we arrive at
\begin{equation}
\label{corfree}
\delta F = \frac{\delta I_E}{\beta}
= - \frac{V_3 V(S^5)}{16 \pi G}
 \frac{\gamma}{h^6}
r_0^4 \left[
  15  + \frac{111}{7} \left( \frac{\eul}{r_0} \right)^2  +
\frac{5885}{96} \left( \frac{\eul}{r_0} \right)^4  + \ldots \right]
\end{equation}
Here we have also used the expansion for the horizon radius
\begin{equation}
r_H = r_0\left[ 1 + \frac{1}{4}  \left( \frac{\eul}{r_0} \right)^2
+ \frac{1}{32} \left( \frac{\eul}{r_0} \right)^4
  + \ldots \right]
\end{equation}
which follows from \eqref{horradius} with one angular momentum turned on only
 and the replacement $l = i \eul$.

Finally, we write the result \eqref{corfree} in terms of the thermodynamic
parameters $T$, $\Omega$ using \eqref{r0exp}, \eqref{lexp} which imply,
\begin{subequations}
\begin{equation}
r_0^4   = (T h^2)^4
\left[ 1 +  \frac{1}{\pi^2} \omega^2
 + \frac{1}{2\pi^4} \omega^4   + \Ord(\omega^6)  \right]
\end{equation}
\begin{equation}
-\left( \frac{\eul}{r_0} \right)^2
= \frac{1}{\pi^2} \omega^2
 +  \frac{1}{2\pi^4} \omega^4  + \Ord(\omega^6)
\end{equation}
\end{subequations}
We also use the relations \eqref{D3scal} that enable to transform
to the field theory parameters. Then we obtain the final form of
the correction
\begin{equation}
\label{deltaF}
 \delta F =
- \frac{\zeta(3) \pi^2}{64} (2\lambda)^{-3/2} N^2 V_3 T^4
\left[ 15  - \frac{6}{7\pi^2} \omega^2
+\frac{28151}{672\pi^4} \omega^4  + \Ord(\omega^6) \right]
\end{equation}
This gives the interpolation function
\begin{equation}
\label{generalf}
f(\lambda,\omega)
= \frac{3}{4} - \frac{3}{8\pi^2} \omega^2 + \frac{69}{64\pi^4}\omega^4
+ \frac{\zeta(3)}{64} (2\lambda)^{-3/2}
\left( 90 - \frac{981}{7\pi^2} \omega^2 + \frac{7655}{16\pi^4} \omega^4 \right)
+ \cdots
\end{equation}
which includes \eqref{flambdainf} and \eqref{zerofcomp} as special
cases.
Because of \eqref{inequal} we expect the correction term in
\eqref{generalf} to be positive for \( \omega < 1 \) and
this is indeed the case.
Since the corrections away from $\lambda=\infty$ behave as expected,
we consider this as further evidence for the
interpolation conjecture \eqref{interD3} in the case of spinning D3-branes,

It is not a priori apparent that the method used to compute
\eqref{generalf} gives the full result to first order in $\gamma$,
since this semi-classical method does not consider induced
perturbations of the geometry. However, we note that for the
non-rotating D3-brane case, the perturbed metric induced by the
correction term \eqref{coract} was shown
\cite{Gubser:1998nz,Pawelczyk:1998pb} to yield the same correction
to the free energy as the semiclassical approximation in which the
correction is evaluated for the original unperturbed
metric\footnote{This feature has been shown to persist as well for
the corrections to the near-horizon M2 and M5-brane backgrounds 
originating from the $R^4$ term in M-theory
\cite{Caldarelli:1999ar}.}. Thus, it seems that the thermodynamics
somehow disregards perturbations of the geometry. We expect this
to hold also for non-dilatonic spinning branes. It would be
interesting, though technically difficult, to find the actual
perturbed metric for the spinning D3-brane and determine whether
the result is still given by \eqref{generalf}. Another interesting
check on the interpolation function \eqref{generalf} would be to
compute the \( 1/\lambda \) correction to the boundary of
stability. From the values in Table \ref{tabcompomega} we would
expect this correction to be positive.


\section*{Acknowledgments}
We thank J. Ambj{\o}rn, E. Cheung, P. Damgaard, P. Di Vecchia, E. Kiritsis, F. Larsen,
R. Marotta,
J. L. Petersen, B. Pioline, K. Savvidis, K. Sfetsos and R. Szabo
for useful discussions and correspondence,
and we thank J. Correia for useful discussions and
early participation on the subjects of Section 2.
NO thanks the
CERN Theory Division for support and hospitality during completion
of this work. This work is supported in part by TMR network ERBFMRXCT96-0045.
\begin{appendix}

\section{Thermodynamics of black p-branes in the near-horizon limit}
\label{appenergy}

In this appendix we consider the thermodynamics, and in particular
the energy above extremality in the near-horizon limit of a
general non-rotating $p$-brane. We also derive the Smarr formula
and check that it is fulfilled. This means that the first law of
thermodynamics is obeyed in the near-horizon limit, and that there
are no extra thermodynamic parameters related to the charge. We
begin by giving a short review of the classification of $p$-branes
that preserve a certain fraction of the  supersymmetry (in the
extremal limit).

A black $p$-brane solution of the action \eqref{pbact} is
characterized by a particular value of $a$. If we define
\begin{equation}
b \equiv \frac{2(D-2)}{(p+1)(d-2)+\frac{1}{2}a^2(D-2)}
\end{equation}
then the non-rotating black $p$-brane background takes the form
\begin{subequations}
\label{pbrasol}
\begin{equation}
ds^2 = H^{-\frac{d-2}{D-2}b} \Big( - f dt^2 + \sum_{i=1}^p (dy^i)^2 \Big)
+ H^{\frac{p+1}{D-2}b} \Big( f^{-1} dr^2 + r^2 d\Omega_{d-1}^2 \Big)
\end{equation}
\begin{equation}
e^\phi = H^{\frac{a}{2}b}
\end{equation}
\begin{equation}
A_{p+1} = (-1)^p \sqrt{b} \coth \alpha \Big( H^{-1} -1 \Big)
dt \wedge dy^1 \wedge dy^2 \wedge \cdots \wedge dy^p
\end{equation}
\end{subequations}
with
\begin{equation}
H = 1 + \frac{r_0^{d-2} \sinh^2 \alpha}{r^{d-2}}
\sp
f = 1 - \frac{r_0^{d-2}}{r^{d-2}}
\end{equation}
This $p$-brane solution is a $1/2^b$-BPS state
\cite{Behrndt:1999mk} for \( r_0 = 0 \), so that the spinning
solutions discussed in the text correspond to $b=1$. Table
\ref{tabbranes} lists the most common branes together with the
corresponding values of $D$, $a$ and $b$ 
(a more extensive list with other values of
$b$ can be found in Ref. \cite{Behrndt:1999mk}).  

\begin{table}
\begin{center}
\begin{tabular}{|c|c|c|c|c|}
\hline Brane  &  Theory  &  $D$  &  $a$  &  $b$  \\ \hline M2   &
M & 11   & 0       &  1 \\ \hline M5   & M & 11   & 0 & 1
\\ \hline D$p$ & string  & 10   & $(3-p)/2$ & 1 \\ \hline
NS1  & string  & 10   & $-1$      &  1 \\ \hline NS5  & string  &
10 & 1 & 1
\\ \hline d$p$ &  little string  & 6    & $1-p$ & 2
\\ \hline
\end{tabular}
\end{center}
\caption{The characteristic numbers $a$ and $b$ for branes with
$b$ equal to 1 or 2.\label{tabbranes}}
\end{table}

The thermodynamic quantities of the background \eqref{pbrasol} are
\begin{subequations}
\label{thermgenb}
\begin{equation}
M = \frac{V_p V(S^{d-1})}{16 \pi G} r_0^{d-2}
\Big[ d-1 + b (d-2) \sinh^2 \alpha \Big]
\end{equation}
\begin{equation}
\label{tseqs}
T = \frac{d-2}{4\pi r_0} (\cosh \alpha)^{-b}
\sp
S = \frac{V_p V(S^{d-1})}{4G} r_0^{d-1} ( \cosh \alpha )^{b}
\end{equation}
\begin{equation}
\mu = \tanh \alpha
\sp
Q = \frac{V_p V(S^{d-1})}{16 \pi G} b (d-2) r_0^{d-2}
\cosh \alpha \sinh \alpha
\end{equation}
\end{subequations}
satisfying the Smarr formula
\begin{equation}
\label{smarrr}
(d-2) M = (d-1) TS + (d-2) \mu Q
\end{equation}
and the first law of thermodynamics
\begin{equation}
dM = T dS + \mu dQ \sp M = M(S,Q)
\end{equation}
The energy above extremality is
\begin{equation}
\label{inten}
E = M - Q = \frac{V_p V(S^{d-1})}{16 \pi G} r_0^{d-2}
\Big[ d-1 + b (d-2) (\sinh^2 \alpha - \cosh \alpha \sinh \alpha ) \Big]
\end{equation}

The near-horizon limit is defined via the rescaling
\begin{subequations}
\label{rescal2}
\begin{equation}
r  = \frac{\old{r}}{\ell^2}   \sp
r_0  = \frac{\oldp{r_0}}{\ell^2}  \sp
h^{d-2} = \frac{\old{h}^{d-2}}{\ell^{2d-4-\frac{4}{b}}}
\end{equation}
\begin{equation}
ds^2 = \frac{\oldp{ds^2}}{\ell^{4(d-2)b/(D-2)}} \sp
e^{\phi} = \ell^{2a} e^{\phi_{\rm old}} \sp
A = \frac{A_{\rm old}}{\ell^4} \sp
G  = \frac{\old{G}}{\ell^{2(d-2)}}
\end{equation}
\end{subequations}
and taking \( \ell \rightarrow 0 \), keeping the old quantities
fixed. Note that the near-horizon limit depends on the fraction of
supersymmetries that is preserved and that we recover
\eqref{rescal} for $b =1$ . Using this limit in \eqref{tseqs} and
\eqref{inten} we obtain
\begin{subequations}
\label{tseqsnh}
\begin{equation}
T = \frac{d-2}{4\pi r_0} \Big( \frac{r_0}{h} \Big)^{\frac{d-2}{2} b}
\sp
S = \frac{V_p V(S^{d-1})}{4G} r_0^{d-1}
\Big( \frac{h}{r_0} \Big)^{\frac{d-2}{2} b}
\end{equation}
\begin{equation}
\label{intennh}
E = \frac{V_p V(S^{d-1})}{16 \pi G} 
\Big[ d-1 - \frac{b}{2} (d-2) \Big]r_0^{d-2}
\end{equation}
\end{subequations}

To derive the Smarr formula we consider  the canonical rescaling
\begin{equation}
h \rightarrow h
\sp
r_0 \rightarrow \lambda r_0
\end{equation}
under which we have the transformation
\begin{equation}
E \rightarrow \lambda^{d-2}
\sp
S \rightarrow \lambda^{(1-\frac{1}{2}b) d +b-1 } S
\end{equation}
This gives the Smarr formula
\begin{equation}
\label{smarrrr}
(d-2) E = \Big( d-1 - \frac{b}{2} (d-2) \Big) TS
\end{equation}
corresponding to the first law of thermodynamics
\begin{equation}
 dE = T dS \sp E=E(S)
\end{equation}
The Smarr formula \eqref{smarrrr} is indeed satisfied with
\eqref{tseqsnh}. We note that \eqref{smarrrr}
is qualitatively different from the asymptotically-flat black
brane Smarr formula \eqref{smarrr} since it exhibits  a dependence
on the amount of unbroken supersymmetry that the brane has in the
extremal limit.

The free energy is given by
\begin{equation}
F = E - TS
= - \frac{\frac{b}{2}(d-2) - 1}{d-1 - \frac{b}{2} (d-2)} E
= -  \frac{V_p V(S^{d-1})}{16 \pi G} \Big[ \frac{b}{2} (d-2) -1 \Big] r_0^{d-2}
\end{equation}
In particular, for $b=1$ we have
\begin{equation}
\label{eandfb1}
E = \frac{V_p V(S^{d-1})}{16 \pi G} \frac{d}{2} r_0^{d-2}
\sp
F = - \frac{d-4}{d} E
= - \frac{V_p V(S^{d-1})}{16 \pi G} \frac{d-4}{2} r_0^{d-2}
\end{equation}
while for $b=2$ the result reads
\begin{equation}
E = \frac{V_p V(S^{d-1})}{16 \pi G} r_0^{d-2}
\sp
F = - (d-3) E
= -\frac{V_p V(S^{d-1})}{16 \pi G} (d-3) r_0^{d-2}
\end{equation}
\section{Spheroidal coordinates \label{sphcoor}}

In this appendix we define the spheroidal coordinates for a
$d$-dimensional Euclidean space with Cartesian coordinates $x^a,
a=1 \ldots d$. We define the metric
\begin{equation}
(ds_d)^2 = \sum_{a=1}^d (dx^a)^2
\end{equation}
and treat the cases $d$ even and odd separately.

\und{The case $d=2n$}

The spheroidal coordinates are the ``radius'' $r$ and the angles
$ \theta , \psi_1 ,..., \psi_{n-2} ,$, $ \phi_1 ,..., \phi_n $.
Define the quantities
\begin{eqnarray}
&& \mu_1 = \sin \theta,\ \ \mu_2 = \cos \theta \sin \psi_1,\ \
\mu_3 = \cos \theta \cos \psi_1 \sin \psi_2 \sp \ldots \sp \nn \\
&& \mu_{n-1} = \cos \theta \cos \psi_1 \cdots \cos \psi_{n-3} \sin
\psi_{n-2},\ \ \mu_n = \cos \theta \cos \psi_1 \cdots \cos
\psi_{n-2}
\end{eqnarray}
which satisfy
\begin{equation}
\sum_{i=1}^n \mu_i^2 =1
\end{equation}
The spheroidal coordinates are then defined by
\begin{equation}
x^{2i-1}  =  \sqrt{r^2+l_i^2} \mu_i \cos \phi_i \sp \sp x^{2i} =
\sqrt{r^2+l_i^2} \mu_i \sin \phi_i \sp i = 1 \ldots n
\end{equation}
The coordinates \( \phi_1,...,\phi_n \) are the rotation angles
and \( l_1,...,l_n \) correspond to the angular momenta in these
angles. We have
\begin{equation}
\sum_{a=1}^d (x^a)^2  = r^2 + \sum_{i=1}^n l_i^2 \mu_i^2
\end{equation}
and the ranges of the angles are given by
\begin{equation}
0 \leq \theta , \psi_1 ,..., \psi_{n-2} \leq \frac{\pi}{2},\ \ 0
\leq \phi_1 ,..., \phi_n \leq 2\pi
\end{equation}

\und{The case $d = 2n+1$}

The spheroidal coordinates are the ``radius'' $r$ and the angles
$ \theta , \psi_1 ,..., \psi_{n-1}$, $\phi_1 ,..., \phi_n $.
Define the quantities
\begin{eqnarray}
&& \mu_1 = \sin \theta,\ \ \mu_2 = \cos \theta \sin \psi_1,\ \
\mu_3 = \cos \theta \cos \psi_1 \sin \psi_2 \sp \ldots \sp \nn \\
&& \mu_n = \cos \theta \cos \psi_1 \cdots \cos \psi_{n-2} \sin
\psi_{n-1},\ \ \mu_{n+1} = \cos \theta \cos \psi_1 \cdots \cos
\psi_{n-1}
\end{eqnarray}
which satisfy
\begin{equation}
\sum_{i=1}^{n+1} \mu_i^2 =1
\end{equation}
The spheroidal coordinates are then defined by
\begin{subequations}
\begin{equation}
x^{2i-1}  =  \sqrt{r^2+l_i^2} \mu_i \cos \phi_i \sp x^{2i} =
\sqrt{r^2+l_i^2} \mu_i \sin \phi_i \sp i = 1 \ldots n
\end{equation}
\begin{equation}
x^d = r \mu_{n+1}
\end{equation}
\end{subequations}
The coordinates \( \phi_1,...,\phi_n \) are the rotation angles
and \( l_1,...,l_n \) correspond to the angular momenta in these
angles. In this case, we have
\begin{equation}
\sum_{a=1}^d (x^a)^2  = r^2 + \sum_{i=1}^n l_i^2 \mu_i^2
\end{equation}
The ranges of the angles are, for \( d \geq 5 \), given by
\begin{equation}
0 \leq \theta , \psi_1 ,..., \psi_{n-2} \leq \frac{\pi}{2},\ \ 0
\leq \psi_{n-1} \leq \pi,\ \ 0 \leq \phi_1 ,..., \phi_n \leq 2\pi
\end{equation}
and for \(d=3\) we have
\begin{equation}
0 \leq \theta \leq \pi,\ \ 0 \leq \phi_1 \leq 2\pi
\end{equation}
%

\und{The spheroidal metric}

The metric in spheroidal coordinates takes the form
\begin{equation}
\sum_{a=1}^d (dx^a)^2 = K_d dr^2 + \Lambda_{\alpha \beta}
d\eta^\alpha d\eta^\beta
\end{equation}
where $\eta^{\alpha}$ denote the set of angular coordinates. For
the radial coordinate  the metric component  takes the form
\begin{equation}
g_{rr} = K_d (r,\theta,\psi_1,...,\psi_{d-n-2}) \equiv \left\{
\begin{array}{ll} \sum_{i=1}^n \mu_i^2 \Big( 1+\frac{l_i^2}{r^2}
\Big)^{-1}
 & \sp  d=2n \\
\sum_{i=1}^n \mu_i^2 \Big( 1+\frac{l_i^2}{r^2} \Big)^{-1} +
\mu_{n+1}^2
 & \sp d=2n+1
\end{array} \right.
\end{equation}
and the general form of the remaining non-zero components is \eqa{
g_{\theta \theta} &=& r^2 + l_1^2 \cos^2 \theta + \tan^2 \theta
\Big( \mu_2^2 l_2^2 + \cdots + \mu_n^2 l_n^2 \Big)
\\
g_{\psi_1 \psi_1} &=& \cos^2 \theta \Big( r^2 + l_2^2 \cos^2
\psi_1 \Big) + \tan^2 \psi_1 \Big( \mu_3^2 l_3^2 + \cdots +
\mu_n^2 l_n^2 \Big)
\\
g_{\psi_2 \psi_2} &=& \cos^2 \theta \cos^2 \psi_1 \Big( r^2 +
l_3^2 \cos^2 \psi_2 \Big) + \tan^2 \psi_2 \Big( \mu_4^2 l_4^2 +
\cdots + \mu_n^2 l_n^2 \Big)
\\
g_{\psi_3 \psi_3} &=& \cos^2 \theta \cos^2 \psi_1 \cos^2 \psi_2
\Big( r^2 + l_4^2 \cos^2 \psi_3 \Big) + \tan^2 \psi_3 \Big(
\mu_5^2 l_5^2 + \cdots + \mu_n^2 l_n^2 \Big)
\\
g_{\psi_{n-2} \psi_{n-2}} &=& \cos^2 \theta \cos^2 \psi_1 \cdots
\cos^2 \psi_{n-3} \Big( r^2 + l_{n-1}^2 \cos^2 \psi_{n-2}
+ l_n^2 \sin^2 \psi_{n-2} \Big)
\\
g_{\theta \psi_1} &=& - \tan \theta \cot \psi_1 \mu_2^2 l_2^2 +
\tan \theta \tan \psi_1 \Big( \mu_3^2 l_3^2 + \cdots + \mu_n^2
l_n^2 \Big)
\\
g_{\theta \psi_2} &=& - \tan \theta \cot \psi_2 \mu_3^2 l_3^2 +
\tan \theta \tan \psi_2 \Big( \mu_4^2 l_4^2 + \cdots + \mu_n^2
l_n^2 \Big)
\\
g_{\psi_1 \psi_2} &=& - \tan \psi_1 \cot \psi_2 \mu_3^2 l_3^2 +
\tan \psi_1 \tan \psi_2 \Big( \mu_4^2 l_4^2 + \cdots + \mu_n^2
l_n^2 \Big)
\\
g_{\phi_i \phi_i} &=& \mu_i^2 (r^2 + l_i^2) \sp  i=1 \ldots n }
for both $d=2n$ and $d=2n+1$. As an aid to the reader we list
below the angles and explicit expressions for the spheroidal
metric when $ 3 \leq d \leq 9$.
\begin{subequations}
\begin{equation}
 d=3 \co  \theta,\phi_1
\end{equation}
\begin{equation} \nn (ds_3)^2 = K_3 dr^2
+ \Big(r^2 + l_1^2 \cos^2 \theta \Big) d\theta^2 + \sin^2 \theta
\Big(r^2 + l_1^2\Big) d\phi_1^2 \end{equation}
\begin{equation}
 d=4 \co  \theta,\phi_1,\phi_2
\end{equation}
\eqa{ && (ds_4)^2 = K_4 dr^2 + \Big( r^2 + l_1^2 \cos^2 \theta +
l_2^2 \sin^2 \theta \Big) d\theta^2 + \sin^2 \theta \Big(r^2
+l_1^2\Big) d\phi_1^2 \hfill
\\ &&
+ \cos^2 \theta \Big(r^2 +l_2^2\Big) d\phi_2^2 }
\begin{equation}
  d=5 \co  \theta, \psi_1,\phi_1,\phi_2
\end{equation}
\eqa{ && (ds_5)^2 = K_5 dr^2 + \Big( r^2 + l_1^2 \cos^2 \theta +
l_2^2 \sin^2 \theta \sin^2 \psi_1 \Big) d\theta^2 + \cos^2 \theta
\Big( r^2 + l_2^2 \cos^2 \psi_1 \Big) d\psi_1^2
\\ &&
- 2 l_2^2 \cos \theta \sin \theta \cos \psi_1 \sin \psi_1 d\theta
d\psi_1 + \sin^2 \theta \Big(r^2+l_1^2\Big) d\phi_1^2 + \cos^2
\theta \sin^2 \psi_1 \Big(r^2+l_2^2\Big) d\phi_2^2 }
\begin{equation}
 d=6 \co \theta, \psi_1,\phi_1,\phi_2,\phi_3
\end{equation}
\eqa{ && (ds_6)^2 = K_6 dr^2 + \Big( r^2 + l_1^2 \cos^2 \theta +
l_2^2 \sin^2 \theta \sin^2 \psi_1 + l_3^2 \sin^2 \theta \cos^2
\psi_1 \Big) d\theta^2
\\ &&
+ \cos^2 \theta \Big( r^2 + l_2^2 \cos^2 \psi_1 + l_3^2 \sin^2
\psi_1 \Big) d\psi_1^2 + 2 \cos \theta \sin \theta \cos \psi_1
\sin \psi_1 \Big( -l_2^2 + l_3^2 \Big) d\theta d\psi_1
\\ &&
+ \sin^2 \theta \Big(r^2 +l_1^2\Big) d\phi_1^2 + \cos^2 \theta
\sin^2 \psi_1 \Big(r^2 +l_2^2\Big) d\phi_2^2 + \cos^2 \theta
\cos^2 \psi_1 \Big(r^2 +l_3^2\Big) d\phi_3^2 }
\begin{equation}
 d=7 \co \theta,\psi_1,\psi_2,\phi_1,\phi_2,\phi_3
\end{equation}
\eqa{ && (ds_7)^2 = K_7 dr^2 + \Big( r^2 + l_1^2 \cos^2 \theta +
l_2^2 \sin^2 \theta \sin^2 \psi_1 + l_3^2 \sin^2 \theta \cos^2
\psi_1 \sin^2 \psi_2 \Big) d\theta^2
\\ &&
+ \cos^2 \theta \Big( r^2 + l_2^2 \cos^2 \psi_1 + l_3^2 \sin^2
\theta \sin^2 \psi_2 \Big) d\psi_1^2 + \cos^2 \theta \cos^2 \psi_1
\Big( r^2 + l_3^2 \cos^2 \psi_2 \Big) d\psi_2^2
\\ &&
+ 2\cos \theta \sin \theta \cos \psi_1 \sin \psi_1 \Big( -l_2^2 +
l_3^2 \sin^2 \psi_2 \Big) d\theta d\psi_1
\\ &&
- 2 \cos \theta \sin \theta \cos^2 \psi_1 \cos \psi_2 \sin \psi_2
l_3^2 d\theta d\psi_2
\\ &&
- 2 l_3^2 \cos \psi_1 \sin \psi_1 \cos^2 \psi_1 \cos \psi_2 \sin
\psi_2 d\psi_1 d\psi_2 + \sin^2 \theta \Big(r^2 +l_1^2\Big)
d\phi_1^2
\\ &&
+ \cos^2 \theta \sin^2 \psi_1 \Big(r^2 +l_2^2\Big) d\phi_2^2 +
\cos^2 \theta \cos^2 \psi_1 \sin^2 \psi_2 \Big(r^2 +l_3^2\Big)
d\phi_3^2 }
\begin{equation}
d=8 \co \theta,\psi_1,\psi_2,\phi_1,\phi_2,\phi_3,\phi_4
\end{equation}
\eqa{ && (ds_8)^2 = K_8 dr^2 + \Big( r^2 + l_1^2 \cos^2 \theta +
l_2^2 \sin^2 \theta \sin^2 \psi_1 + l_3^2 \sin^2 \theta \cos^2
\psi_1 \sin^2 \psi_2
\\ &&
+ l_4^2 \sin^2 \theta \cos^2 \psi_1 \cos^2 \psi_2 \Big) d\theta^2
\\ &&
+ \cos^2 \theta \Big( r^2 + l_2^2 \cos^2 \psi_1 + l_3^2 \sin^2
\psi_1 \sin^2 \psi_2 + l_4^2 \sin^2 \psi_1 \cos^2 \psi_2 \Big)
d\psi_1^2
\\ &&
+ \cos^2 \theta \cos^2 \psi_1 \Big( r^2 + l_3^2 \cos^2 \psi_2 +
l_4^2 \sin^2 \psi_2 \Big) d\psi_2^2
\\ &&
+ 2 \cos \theta \sin \theta \cos \psi_1 \sin \psi_1 \Big( -l_2^2 +
l_3^2 \sin^2 \psi_2 + l_4^2 \cos^2 \psi_2 \Big) d\theta d\psi_1
\\ &&
+ 2 \cos \theta \sin \theta \cos^2 \psi_1 \cos \psi_2 \sin \psi_2
\Big( -l_3^2 + l_4^2 \Big) d\theta d\psi_2
\\ &&
+ 2 \cos^2 \theta \cos \psi_1 \sin \psi_1 \cos \psi_2 \sin \psi_2
\Big( - l_3^2 + l_4^2 \Big) d\psi_1 d\psi_2 + \sin^2 \theta
\Big(r^2 +l_1^2 \Big) d\phi_1^2
\\ &&
+ \cos^2 \theta \sin^2 \psi_1 \Big(r^2 +l_2^2 \Big) d\phi_2^2 +
\cos^2 \theta \cos^2 \psi_1 \sin^2 \psi_2 \Big(r^2 +l_3^2 \Big)
d\phi_3^2
\\ &&
+ \cos^2 \theta \cos^2 \psi_1 \cos^2 \psi_2 \Big(r^2 +l_4^2 \Big)
d\phi_4^2 }
\begin{equation}
 d=9 \co \theta,\psi_1,\psi_2,\psi_3,\phi_1,\phi_2,\phi_3,\phi_4
\end{equation}
\eqa{ && (ds_9)^2 = K_9 dr^2 + \Big( r^2 + l_1^2 \cos^2 \theta +
l_2^2 \sin^2 \theta \sin^2 \psi_1 + l_3^2 \sin^2 \theta \cos^2
\psi_1 \sin^2 \psi_2
\\ &&
+ l_4^2 \sin^2 \theta \cos^2 \psi_1 \cos^2 \psi_2 \sin^2 \psi_3
\Big) d \theta^2 + \cos^2 \theta \Big( r^2 + l_2^2 \cos^2 \psi_1 +
l_3^2 \sin^2 \psi_1 \sin^2 \psi_2
\\ &&
+ l_4^2 \sin^2 \psi_1 \cos^2 \psi_2 \sin^2 \psi_3 \Big) d \psi_1^2
+ \cos^2 \theta \cos^2 \psi_1 \Big( r^2 + l_3^2 \cos^2 \psi_2 +
l_4^2 \sin^2 \psi_2 \sin^2 \psi_3 \Big) d\psi_2^2
\\ &&
+ \cos^2 \theta \cos^2 \psi_1 \cos^2 \psi_2 \Big( r^2 + l_4^2
\cos^2 \psi_3 \Big) d\psi_3^2
\\ &&
+ 2 \cos \theta \sin \theta \cos \psi_1 \sin \psi_1 \Big( -l_2^2 +
l_3^2 \sin^2 \psi_2 + l_4^2 \cos^2 \psi_2 \sin^2 \psi_3 \Big) d
\theta d \psi_1
\\ &&
+ 2 \cos \theta \sin \theta \cos^2 \psi_1 \cos \psi_2 \sin \psi_2
\Big( -l_3^2 + l_4^2 \sin^2 \psi_3 \Big) d \theta d \psi_2
\\ &&
- 2 l_4^2 \cos \theta \sin \theta \cos^2 \psi_1 \cos^2 \psi_2 \cos
\psi_3 \sin \psi_3 d \theta d \psi_3
\\ &&
+ 2\cos^2 \theta \cos \psi_1 \sin \psi_1 \cos \psi_2 \sin \psi_2
\Big( -l_3^2 + l_4^2 \sin^2 \psi_3 \Big) d \psi_1 d \psi_2
\\ &&
- 2 l_4^2 \cos^2 \theta \cos^2 \psi_1 \cos \psi_2 \sin \psi_2 \cos
\psi_3 \sin \psi_3 d \psi_2 d \psi_3 + \sin^2 \theta \Big(r^2
+l_1^2 \Big) d\phi_1^2
\\ &&
+ \cos^2 \theta \sin^2 \psi_1 \Big(r^2 +l_2^2 \Big) d\phi_2^2 +
\cos^2 \theta \cos^2 \psi_1 \sin^2 \psi_2 \Big(r^2 +l_3^2 \Big)
d\phi_3^2
\\ &&
+ \cos^2 \theta \cos^2 \psi_1 \cos^2 \psi_2 \sin^2 \psi_3 \Big(r^2
+l_4^2 \Big) d\phi_4^2 }
\end{subequations}

\und{Spheroidal metric with one angular momentum}

As an aid to the reader we also give the explicit expressions for
the spheroidal metric when only one angular momentum \( l_1 = l \)
is non-zero,
\begin{subequations}
\begin{equation}
(ds_3)^2 = \Big( 1 - \frac{l^2 \sin^2 \theta }{l^2 + r^2 } \Big)
dr^2 + \Big(r^2 + l^2 \cos^2 \theta \Big) d\theta^2 + \sin^2
\theta \Big(r^2 + l^2\Big) d\phi_1^2
\end{equation}
\begin{equation}
(ds_4)^2 = \Big( 1 - \frac{l^2 \sin^2 \theta }{l^2 + r^2 } \Big)
dr^2 + \Big( r^2 + l^2 \cos^2 \theta \Big) d\theta^2 + \sin^2
\theta \Big(r^2 +l^2\Big) d\phi_1^2 + r^2 \cos^2 \theta d\phi_2^2
\end{equation}
\begin{eqnarray}
&& (ds_5)^2 = \Big( 1 - \frac{l^2 \sin^2 \theta }{l^2 + r^2 }
\Big) dr^2 + \Big( r^2 + l^2 \cos^2 \theta \Big) d\theta^2 + r^2
\cos^2 \theta d\psi_1^2 + \sin^2 \theta \Big(r^2+l^2\Big)
d\phi_1^2
 \nn \\ &&
+ r^2 \cos^2 \theta \sin^2 \psi_1 d\phi_2^2
\end{eqnarray}
\begin{eqnarray}
 && (ds_6)^2 = \Big( 1 - \frac{l^2 \sin^2 \theta }{l^2 + r^2 }
\Big) dr^2 + \Big( r^2 + l^2 \cos^2 \theta \Big) d\theta^2 + r^2
\cos^2 \theta d\psi_1^2 + \sin^2 \theta \Big(r^2 +l^2\Big)
d\phi_1^2 \nn \\ && + r^2 \cos^2 \theta \sin^2 \psi_1 d\phi_2^2 +
r^2 \cos^2 \theta \cos^2 \psi_1 d\phi_3^2
\end{eqnarray}%
\begin{eqnarray}
 && (ds_7)^2 = \Big( 1 - \frac{l^2 \sin^2 \theta }{l^2 + r^2
} \Big) dr^2 + \Big( r^2 + l^2 \cos^2 \theta \Big) d\theta^2 + r^2
\cos^2 \theta d\psi_1^2 + r^2 \cos^2 \theta \cos^2 \psi_1
d\psi_2^2 \nn \\ && + \sin^2 \theta \Big(r^2 +l^2\Big) d\phi_1^2 +
r^2 \cos^2 \theta \sin^2 \psi_1 d\phi_2^2 + r^2 \cos^2 \theta
\cos^2 \psi_1 \sin^2 \psi_2 d\phi_3^2
\end{eqnarray}%
\begin{eqnarray}
 && (ds_8)^2 = \Big( 1 - \frac{l^2 \sin^2 \theta }{l^2 + r^2 }
\Big) dr^2 + \Big( r^2 + l^2 \cos^2 \theta \Big) d\theta^2 + r^2
\cos^2 \theta d\psi_1^2 + r^2 \cos^2 \theta \cos^2 \psi_1
d\psi_2^2 \nn \\ && + \sin^2 \theta \Big(r^2 +l^2 \Big) d\phi_1^2
+ r^2 \cos^2 \theta \sin^2 \psi_1 d\phi_2^2 + r^2 \cos^2 \theta
\cos^2 \psi_1 \sin^2 \psi_2 d\phi_3^2 \nn \\ && + r^2 \cos^2
\theta \cos^2 \psi_1 \cos^2 \psi_2 d\phi_4^2
\end{eqnarray}
\begin{eqnarray} && (ds_9)^2 = \Big( 1 - \frac{l^2 \sin^2
\theta }{l^2 + r^2 } \Big) dr^2 + \Big( r^2 + l^2 \cos^2 \theta
\Big) d \theta^2 + r^2 \cos^2 \theta d \psi_1^2 + r^2 \cos^2
\theta \cos^2 \psi_1 d\psi_2^2 \nn \\ && + r^2 \cos^2 \theta
\cos^2 \psi_1 \cos^2 \psi_2 d\psi_3^2 + \sin^2 \theta \Big(r^2
+l^2 \Big) d\phi_1^2 + r^2 \cos^2 \theta \sin^2 \psi_1 d\phi_2^2
\nn \\ && + r^2 \cos^2 \theta \cos^2 \psi_1 \sin^2 \psi_2
d\phi_3^2 + r^2 \cos^2 \theta \cos^2 \psi_1 \cos^2 \psi_2 \sin^2
\psi_3 d\phi_4^2
\end{eqnarray}
\end{subequations}

\section{The Euclidean near-horizon solution
\label{appeuclnhsol} }

In this appendix we give the Euclidean version of the
near-horizon solution \eqref{nhsol} for
one non-zero angular momentum. This comes into play
in Sections
\ref{secfreeenergy}
 and  \ref{secd3corr}
when calculating the value of the action and of the corrected action.
The Euclidean solutions can simply be obtained by performing the Wick rotation
\begin{equation}
\tau = it,\ \  \eul_i = -i l_i
\end{equation}
This induces the replacement $l_i^2 \rightarrow -\eul_i^2$
in the definitions of $L_d$ of \eqref{Ldeq}
and $K_d$, $\Lambda_{\alpha \beta}$  of the spheroidal metric
\eqref{flattr}. In addition, we find that in the metric \eqref{nearmet}
we have $ -f d t^2 \rightarrow f d\tau^2$ as well as
$ l_i d t d\phi_i \rightarrow \eul_i d \tau d \phi_i$
in the off-diagonal terms.
Finally, in the electric $(p+1)$-form potential we replace
\begin{equation}
\Big( H^{-1} d\tau + \frac{r_0^{\frac{d-2}{2}}}{h^{\frac{d-2}{2}}}
\sum_{i=1}^n l_i \mu_i^2 d\phi_i \Big)
\rightarrow
 -i
\Big( H^{-1} d\tau - \frac{r_0^{\frac{d-2}{2}}}{h^{\frac{d-2}{2}}}
\sum_{i=1}^n \eul_i \mu_i^2 d\phi_i \Big)
\end{equation}
The above substitution rules should enable the reader to easily write down
the general Euclidean case, and we confine ourselves with the explicit
form for the case of one angular momentum $\eul \equiv \eul_1 \neq 0$ only
\begin{subequations}
\label{nheucl}
\begin{eqnarray}
\label{euclmet}
ds^2 &=& H^{-\frac{d-2}{D-2}}
\Big( f d\tau^2
+ \sum_{i=1}^p (dy^i)^2 \Big)
+ H^{\frac{p+1}{D-2}} \Big( \bar{f}^{-1} \frac{r^2 -\eul^2 \cos^2 \theta}{r^2 -\eul^2}  dr^2
+ \Lambda_{\alpha \beta} d\eta^\alpha d\eta^\beta \Big)
\nn  \\ &&
- 2 H^{-\frac{d-2}{D-2}}
\frac{1}{ 1 - \frac{\eul^2 \cos^2 \theta}{r^2} } \frac{h^{\frac{d-2}{2}}
r_0^{\frac{d-2}{2}}}{r^{d-2}}
 \eul \sin^2 \theta  d\tau d\phi_1
\end{eqnarray}
\begin{equation}
\label{eucldil}
e^\phi = H^{\frac{a}{2}}
\end{equation}
\begin{equation}
A_{p+1}  = -i (-1)^p
\Big( H^{-1} d \tau  - \frac{r_0^{\frac{d-2}{2}}}{h^{\frac{d-2}{2}}}
\eul \sin^2 \theta d\phi_1 \Big)
\wedge dy^1 \wedge dy^2 \wedge \cdots \wedge dy^p
\end{equation}
\end{subequations}
where
\begin{equation}
H =
\frac{1}{ 1 - \frac{\eul^2 \cos^2 \theta}{r^2} }  \frac{h^{d-2}}{r^{d-2} }
\sp
f = 1 - \frac{1}{ 1 - \frac{\eul^2 \cos^2 \theta}{r^2} } \frac{r_0^{d-2}}{r^{d-2} }\sp
\bar f = 1 - \frac{1}{ 1 - \frac{\eul^2 }{r^2} } \frac{r_0^{d-2}}{r^{d-2} }
\end{equation}
and the expressions for $\Lambda_{\alpha \beta}$ in the one-angular momentum
case can be found in Appendix \ref{sphcoor}.

\section{Change of variables from $(r_0,\{l_i\})$ to $(T,\{ \Omega_i \})$
\label{basisch}}

In this appendix we give the formulae needed to go from the
supergravity variables $(r_0,\{l_i\})$ to the thermodynamic
quantities $(T,\{ \Omega_i \})$. Since it  is not possible to
obtain closed expressions (for general $d$) for this change of
variables, we perform this analysis in a weak angular momentum
expansion
\begin{equation}
\frac{l_i}{r_0} \ll  1
\end{equation}
keeping the first three terms only, which suffices for the
applications of the text.

We use  expressions \eqref{limthermo} for $(T,\{\Omega_i\})$,
\begin{equation}
\label{TOrel} T = \frac{d-2-2\kappa}{4 \pi r_H}
\frac{r_0^{\frac{d-2}{2}}}{h^{\frac{d-2}{2}}} \sp \Omega_i =
\frac{l_i}{(l_i^2 +r_H^2)}
\frac{r_0^{\frac{d-2}{2}}}{h^{\frac{d-2}{2}}}
\end{equation}
 to compute the quantities $(r_0,\{l_i\})$ in terms of
the former. For this we first need to use the relation
\eqref{horizon} determining the horizon radius $r_H$ in terms of
these, and we find
\begin{multline}
\label{horradius} r_H = r_0\left[ 1 - \frac{1}{d-2} \sum_i \left(
\frac{l_i}{r_0} \right)^2 - \frac{3}{2(d-2)^2} \left( \sum_i
\left( \frac{l_i}{r_0} \right)^2 \right)^2 \right. \\ \left. +
\frac{1}{2(d-2)} \sum_i \left( \frac{l_i}{r_0} \right)^4
  + \ldots \right]
\end{multline}
Substituting this in \eqref{TOrel} we obtain the expressions
\begin{subequations}
\begin{multline}
T= \frac{d-2}{4 \pi} \frac{r_0^{(d-4)/2}}{h^{(d-2)/2}} \left[ 1
 -\frac{1}{d-2} \sum_i \left( \frac{l_i}{r_0} \right)^2
- \frac{7}{2(d-2)^2} \left( \sum_i \left( \frac{l_i }{r_0}
\right)^2 \right)^2 \right. \\ \left. + \frac{3}{2(d-2)} \sum_i
\left( \frac{l_i}{r_0} \right)^4 +\ldots \right]
\end{multline}
\begin{equation}
\Omega_i^2 = \frac{r_0^{d-4}}{h^{d-2}}
  \left( \frac{l_i}{r_0} \right)^2
\left[ 1 + \frac{4}{d-2} \sum_j \left( \frac{l_j}{r_0} \right)^2
-2  \left( \frac{l_i}{r_0} \right)^2 +\ldots \right]
\end{equation}
\end{subequations}
which can be inverted to give
\begin{subequations}
\begin{multline}
r_0 = \left(  \tilde T h^{(d-2)/2}  \right)^{2/(d-4)} \left[ 1 +
\frac{2}{(d-4)(d-2)} \sum_i \tilde \omega_i^2 \right.
\\
 \left. - \frac{2(2d-9)}{(d-2)^2(d-4)^2}
\left( \sum_i \tilde \omega_i^2
\right)^2
 +  \frac{1}{(d-2)(d-4)}
\sum_i \tilde \omega_i^4  + \ldots
\right]
\end{multline}
\begin{equation}
\label{lexp} \left( \frac{l_i}{r_0} \right)^2
=
 \tilde \omega_i^2 \left[ 1 -
  \frac{6}{d-2} \sum_j \tilde \omega_j^2
  +2
   \tilde \omega_i^2
+ \ldots \right]
\end{equation}
\end{subequations}
where we have defined
\begin{equation}
\tilde T \equiv \frac{4 \pi T}{d-2}
\sp \tilde \omega_i =
\frac{\Omega_i}{\tilde T}
\end{equation}
Finally, we also give the expression
\begin{multline}
\label{r0exp} r_0^{d-2} = \left( \tilde T h^{(d-2)/2}
\right)^{2(d-2)/(d-4)} \left[ 1 + \frac{2}{d-4} \sum_i
 \tilde \omega_i^2 \right.
\\
\left. -   \frac{2(d-6)}{(d-2)(d-4)^2} \left( \sum_i
\tilde \omega_i^2 \right)^2 + \frac{1}{d-4}
\sum_i \tilde \omega_i^4  + \ldots
\right]
\end{multline}
which enters the free energy \eqref{nhgibbs}.

\section{Polylogarithms \label{apppoly} }

\newcommand{\re}{\mathrm{Re}}
\newcommand{\im}{\mathrm{Im}}

In this appendix we define the polylogarithm functions and give some general
properties. We also discuss a continuation of the polylogarithms to
real numbers greater than one, which is used in Section \ref{subsecweakfe}.
The $n$th polylogarithm function is defined as
\begin{equation}
\li_n (z) = \sum_{k=1}^\infty \frac{z^k}{k^n}
\end{equation}
for \( z \in \mathbb{C} - \{ u \in \mathbb{R}+ 2\pi i \mathbb{Z} |
\re(u) > 1 \} \), where \( \re(u) \) means the real part of \( u
\).
satisfying
\begin{equation}
\li_n ( 1 ) = \zeta(n)\  \mathrm{for}\ n \neq 1
\end{equation}
We also have the relation
\begin{equation}
\li_n(-1) = \tilde{\zeta}(n)
\end{equation}
where we have defined
\begin{equation}
\tilde{\zeta} (n) = \left\{ \begin{array}{ll} (1-2^{1-n})\zeta(n)\
&  \sp n \neq 1 \\ -\log(2) & \sp n = 1
\end{array} \right.
\end{equation}
The polylogarithm satisfies the integral formula
\begin{equation}
\label{liint} \int_0^\infty dx\, x^{n-2} \log( 1 - e^{z-x} ) = -
\Gamma(n-1) \li_n (e^z)
\end{equation}
for \( z \in \{ u \in \mathbb{C} | \im(u) \not\in 2\pi \mathbb{Z}
\} \).

We also define
\begin{equation}
B_n(z) = \frac{1}{2} \left( \li_n(e^z) + \li_n(e^{-z}) \right)
\end{equation}
for \( z \in \{ u \in \mathbb{C} | \im(u) \in [-\pi,\pi ]-\{0\} \}
\), and
\begin{equation}
F_n(z) = \frac{1}{2} \left( \li_n(-e^z) + \li_n(-e^{-z}) \right)
\end{equation}
for \( z \in \{ u \in \mathbb{C} | \im(u) \in (-\pi,\pi ) \} \).

For even $n$ we have
\begin{equation}
B_n(z) = \sum_{k=0}^{n/2} \zeta(n-2k) \frac{z^{2k}}{(2k)!} \pm
\frac{i \pi}{2} \frac{z^{n-1}}{(n-1)!}
\end{equation}
where the optional sign is the sign of \( \im(z) \). For odd $n$
we have
\begin{eqnarray}
B_n(z) &=& \sum_{k=0}^{\frac{n-3}{2}} \zeta( n-2k )
\frac{z^{2k}}{(2k)!} + \left( \pm \frac{i \pi}{2} +
\sum_{k=1}^{n-1} \frac{1}{k} - \log(z) \right)
\frac{z^{n-1}}{(n-1)!} \nn \\ && + \sum_{k=\frac{n+1}{2}}^{\infty}
\zeta( n-2k ) \frac{z^{2k}}{(2k)!}
\end{eqnarray}
where the optional sign is the sign of \( \im(z) \). These
functions satisfy
\begin{equation}
B_{n-2} (z) = \frac{d^2}{dz^2} B_n (z)
\end{equation}
for any $n$. Considering $F_n$, we find that for even $n$ we have
\begin{equation}
F_n(z) = \sum_{k=0}^{n/2} \tilde{\zeta}(n-2k) \frac{z^{2k}}{(2k)!}
\end{equation}
and for odd $n$ we have
\begin{equation}
F_n(z) = \sum_{k=0}^{\infty} \tilde{\zeta}(n-2k)
\frac{z^{2k}}{(2k)!}
\end{equation}
It is easy to see that
\begin{equation}
F_{n-2} (z) = \frac{d^2}{dz^2} F_n (z)
\end{equation}
for any $n$.

If we want to define \( B_n(z) \) also for \( z \in \mathbb{R} \)
we can note that while the imaginary part of \( B_n(z) \) changes
sign when crossing the real line, the real part is continuous.
Thus, it is natural to define \( B_n \) on the real line as the
limit of the real part of \( B_n \). This is also what the
principal value prescription gives, since this is \( \frac{1}{2} (
B(x+i\epsilon) + B(x-i\epsilon) ) \) for \( \epsilon \rightarrow
0^+ \) with \( x \in \mathbb{R} \). Thus, let \( x \in \mathbb{R}
\), then for even $n$ we write
\begin{equation}
B_n(x) = \sum_{k=0}^{n/2} \zeta(n-2k) \frac{x^{2k}}{(2k)!}
\end{equation}
and for odd $n$ we write
\begin{equation}
B_n(x) = \sum_{k=0}^{\frac{n-3}{2}} \zeta( n-2k )
\frac{x^{2k}}{(2k)!} + \left( \sum_{k=1}^{n-1} \frac{1}{k} -
\log(x) \right) \frac{x^{n-1}}{(n-1)!} +
\sum_{k=\frac{n+1}{2}}^{\infty} \zeta( n-2k ) \frac{x^{2k}}{(2k)!}
\end{equation}
Again, these functions satisfy
\begin{equation}
B_{n-2} (x) = \frac{d^2}{dx^2} B_n (x)
\end{equation}
for any $n$.

\end{appendix}

\addcontentsline{toc}{section}{References}

\providecommand{\href}[2]{#2}\begingroup\raggedright\endgroup
\end{document}